\documentclass[lettersize,journal]{IEEEtran}
\usepackage[switch]{lineno}
\usepackage{graphicx} 
\usepackage{hyperref}   
\usepackage{amsmath,amsfonts}
\usepackage{array}
\usepackage{textcomp}
\usepackage{stfloats}
\usepackage{url}
\usepackage{verbatim} 
\usepackage{amssymb}
\usepackage{caption}
\usepackage{xcolor}
\usepackage{float} 
\hypersetup{
	colorlinks=true,
	linkcolor=blue,
	citecolor=blue,
	urlcolor=blue,
}
\usepackage{bm}
\usepackage[linesnumbered,ruled]{algorithm2e}
\usepackage{subcaption}

\SetCommentSty{mycommfont}
\begin{document}
\title{A Model-Free Terminal Iterative Learning Control Scheme for Multi-Layer Printing Alignment Control Problems}

\author{Zifeng Wang, Xiaoning Jin, ~\IEEEmembership{Member,~IEEE,}
\thanks{This work was supported in part by the National Science Foundation under Grants CMMI-1943801 and CMMI-1907250. \it{(Corresponding author: Zifeng Wang.)}}
\thanks{The authors are with the Department of Mechanical and Industrial Engineering, Northeastern University, Boston, MA 02115 USA (email: wang.zifen@northeastern.edu; xi.jin@northeastern.edu).}
}

\maketitle

\begin{abstract}
Roll-to-Roll (R2R) printing technologies are promising for high-volume continuous production of substrate-based electronic products. One of the major challenges in R2R flexible electronics printing is achieving tight alignment tolerances, as specified by the device resolution (usually at the micro-meter level), for multi-layer printed electronics. The alignment of the printed patterns in different layers is known as registration. Conventional registration control methods rely on real-time feedback controllers, such as PID control, to regulate the web tension and the web speed. However, those methods may lose effectiveness in compensating for recurring disturbances and supporting effective mitigation of registration errors. In this paper, we propose a Spatial-Terminal Iterative Learning Control (STILC) method integrated with PID control to iteratively learn and reduce registration error cycle-by-cycle, converging it to zero. This approach enables unprecedented precision in the creation, integration, and manipulation of multi-layer microstructures in R2R processes. We theoretically prove the convergence of the proposed STILC-PID hybrid approach and validate its effectiveness through a simulated registration error scenario caused by axis mismatch between roller and motor, a common issue in R2R systems. The results demonstrate that the STILC-PID hybrid control method can fully eliminate the registration error after a feasible number of iterations. Additionally, we analyze the impact of different learning gains on the convergence performance of STILC.
\end{abstract}

\def\abstractname{Note to Practitioners}
\begin{abstract}
R2R RE control is challenging because the flexible web is susceptible to even tiny disturbances. While standard feedback methods may struggle with the transient and periodic disturbances, a feedforward control scheme like Terminal Iterative Learning Control (TILC) can outperform them by compensating for periodic disturbances with predefined basis functions. Furthermore, we developed STILC to address angle-periodic disturbances caused by axis mismatch, which are common in any rotary machine system. Our primary contribution is a model-free and computationally efficient STILC framework that allows practitioners to design the controller by simply observing the error profile. To ensure successful implementation, we also provide clear design criteria, supported by a rigorous mathematical proof of convergence. We verified the approach through simulation of a common axis mismatch scenario, and the results show that our hybrid STILC-PID method can completely eliminate the recurring registration error within a feasible number of cycles.
\end{abstract}


\section{Introduction}
\label{sec:1}

Roll-to-roll (R2R) printing systems offer a promising approach for high-throughput and continuous manufacturing of substrate-based products, such as thin-film printing electronic devices, batteries, and organic photovoltaics \cite{chen2019towards,kwon2018development,wood2020perspectives,di2018large,salari2019investigation}. R2R processing shows promising industrial scalability for next-generation energy technologies, enabling continuous mass production of both polymer-based energy storage materials and perovskite solar devices \cite{yang2024roll,parvazian2024roll}. However, a critical challenge is inter-layer misalignment, where the overlay of printed patterns across different layers must be precisely controlled to ensure product functionality. This control of registration error (RE) is difficult due to the dynamics of the flexible substrate, or {\it web}.

While RE occurs in both machine and cross-machine directions, longitudinal RE presents the greatest control challenge and is the focus of this work \cite{valimaki2022accuracy}. Longitudinal RE is primarily caused by fluctuations in web tension and speed \cite{chen2020fully,chen2018modeling,kang2013modeling}. The stretchable nature of polymer webs (e.g., PET, PEN) and their low inertia make them susceptible to even tiny disturbances. These control challenges limit the industrial application of R2R printing for many emerging flexible electronics.

Several studies have focused on designing feedback controllers to regulate web tensions and mitigate registration errors. \cite{CHEN2023651} presented a robust linear parameter-varying model predictive control (LPV-MPC) scheme that enhances tension tracking performance by addressing disturbances caused by model uncertainties and slowly-changing dynamics. A data-driven model predictive control (DD-MPC) method was proposed to minimize multistage RE by obtaining the plant model from sensor data and handling multi-input and multi-output systems \cite{shah2022data}. In \cite{pagilla2006decentralized}, a decentralized controller for web processing lines achieves exponential convergence of web tension and transport velocity variations by dividing the system into tension zones and computing equilibrium inputs and reference velocities. A model-based speed variation compensation PD control method has also been developed to eliminate disturbances caused by tension variation during the speed-up phase \cite{zhou2014model}. Furthermore, a fully decoupled proportional-derivative (FDPD) control algorithm was proposed to remove couplings between upstream register control and downstream RE, controlling multiple printing units and accommodating different web lengths between adjacent gravure cylinders \cite{chen2020fully}.

Compared to the aforementioned feedback control methods, feedforward control approaches can provide better system stability when tackling transient disturbances, which is critical when controlling flexible and lightweight objects such as the web in R2R systems. Feedforward controllers generate compensation signals based on their internal settings without waiting for the occurrence of output errors. For instance, Chen et al. developed a model-based feedforward PD (MFPD) control method to reduce the effects of interaction in adjacent print units in an R2R web printing system \cite{chen2016model}. Moreover, angle-periodic disturbances commonly occur in R2R systems, largely due to the rotational nature of mulitple rollers involved in the process. To address these disturbances, researchers have been exploring the use of iterative learning control (ILC), as a type of feedforward control methods for improving web tension control in R2R processes.

Iterative learning control (ILC) is a feedforward technique well-suited for repetitive processes, where the control input for the current cycle is updated based on data from previous cycles \cite{bristow2006survey,lee2007iterative}. Norm Optimal ILC (NOILC), for instance, has successfully regulated web tension by iteratively refining the control profile \cite{sutanto2013norm,sutanto2014vision}. Despite its effectiveness, NOILC has two key limitations in practice:

(1) Model Dependency: NOILC requires an accurate nominal model of the R2R system, which is often difficult to obtain due to nonlinearities and uncertainties.

(2) Restrictive Objective: The control goal is set to strictly track a preset tension profile. In practice, applications often allow for slight deviations in tensions throughout the cycle, while focusing on closely monitoring the RE, which is the {\it terminal output} (i.e., the measurable output at the end of each operation cycle). This strict tracking objective can lead to slow convergence and high computational costs.

To overcome these two limitations of applying ILC in R2R registration control, we propose a Terminal ILC (TILC) framework that utilizes only the terminal measurement of RE from the previous iteration, rather than a tension profile, to update the control input profile. In a multi-layer printing process, the RE can only be measured after the downstream pattern has been printed \cite{kim2017measurement,lee2015register,lee2018smearing}, and the printing of the downstream pattern typically signifies at the end of an operation cycle. Therefore, TILC is a suitable framework for R2R RE control tasks. To eliminate the need for accurate nominal models, we are inspired by earlier ILC research where the simple but effective P-type ILC updating law was prevalent \cite{arimoto1984bettering,4048052}. The P-type ILC updating law can be effectively incorporated into the TILC framework with a predefined basis function. The ideal basis function is designed to provide an input signal profile that compensates for the disturbance in an exactly opposite manner. However, if the basis function deviates significantly from the desired profile, the TILC method may suffer from instabilities, slower convergence, or even failure. Therefore, constructing an appropriate basis function for TILC is crucial to achieving optimal control performance. For instance, a previous study \cite{han2018terminal} employed constant-value basis functions to simplify controller design. That is, the control input remains constant during each iteration, changing only between iterations. However, any system behaviors inside an iteration were ignored, which could have been captured by leveraging prior system knowledge reasonably. A previous study demonstrated that an adaptive basis function obtained by solving the inverse dynamics of the R2R system can successfully make the RE converge to zero \cite{wang2024adaptive,wang2025adaptive}. Other approaches have designed basis functions based on a state-space model of a Rapid Thermal Processing Chemical Vapor Deposition (RTPCVD) system \cite{xu1999terminal} or used data-driven techniques such as iterative dynamical linearization \cite{bu2020data,chi2012data} and neural network-based methods \cite{han2018terminal}. However, these approaches still rely on nominal models identified from online process data, which can be computationally expensive or technically challenging for industrial users.

In this study, we propose a streamlined and effective method for designing TILC controllers to mitigate RE in R2R printing processes, while minimizing hardware computation demand. Observing that perfect tension tracking of the reference profile is not required to achieve zero final RE measurement, we hypothesize that multiple control input profiles, each approximating the reference, can still result in zero RE by the end of the iteration. A P-type TILC updating law can converge to one of these control input profiles, iteratively guaranteeing that the RE approaches zero. Consequently, different from the works in \cite{wang2024adaptive,wang2025adaptive} requiring a well-established nominal model as the prior knowledge, the basis function in this paper can be designed solely by observing tension fluctuations, eliminating the need for prior knowledge or system identification of the R2R system. In this paper, we analyze the convergence conditions theoretically and provide clear guidance for practitioners in designing the TILC controller.

We further refine the basis function design by exploiting the spatial (angular) periodicity inherent in the rollers of R2R systems. Disturbances in such rotary machinery are often dependent on angular displacement rather than time \cite{xu2008spatial}. This motivates the use of Spatial Iterative Learning Control (SILC), which replaces the time index with a spatial one. STILC has been successfully demonstrated in various applications such as additive manufacturing, wind turbines, and robotics. In particular, Afkhami \textit{et al.} designed a SILC method to adjust the voltage pulse width of electrohydrodynamic jet (e-jet) printing processes in a 2D spatial domain based on the layer-wise repeated nature of e-jet printing pocesses \cite{AfkhamiZahra2021SILC,afkhami2023robust,afkhami2021higher}. Liu \textit{et al.} investigated a PD-type SILC for enhancing wind turbine efficiency by adjusting the pitch angle to maintain consistent output power at higher wind speeds, considering that the wind turbine system is spatially periodic in terms of angular displacement \cite{LiuYan2018SILC}. Yang \textit{et al.} presented a SILC method enabling a robot to learn desired paths in unknown environments with fixed spatial constraints by updating its trajectory based on interaction forces that are not periodic in time \cite{yang2022spatial}. Li \textit{et al.} developed an SILC strategy for five-axis CNC machine tools that reduces contour errors by compensating the geometric reference path rather than modifying the controller, demonstrating superior performance compared to traditional tracking error control methods \cite{li2022five}. High-speed trains, with their space-dependent parameters and uncertainties, are another typical class of applications for SILC. It aims to enhance train performance, particularly in terms of tracking speed profiles in the spatial domain \cite{li2021constrained,li2020spatial,zhu2023spatial,huang2023spatial,xin2023spatial}. Kim \textit{et al.} introduced a backlash control algorithm using TILC to mitigate backlash impact in vehicle systems through angle domain control \cite{kim2023terminal}. They designed ILC in the angle domain with access only to the terminal output for a spatial period, which was similar to the strategy we apply to R2R printing processes in this work.

In this study, we propose Spatial-Terminal Iterative Learning Control (STILC), which integrates a spatially-dependent basis function with a decentralized PID controller \cite{pagilla2006decentralized}. This paper extends the initial findings in \cite{wang2023spatial} by providing a rigorous convergence analysis and practical design guidance. The main contributions are:

(1) A model-free STILC framework that mitigates RE without requiring expert knowledge or significant computational resources, offering an alternative to model-based NOILC.

(2) A computationally efficient control update law where the entire input profile is adjusted via a single parameter learned from the terminal RE measurement.

(3) Clear criteria for designing the basis function and learning gain, enabling practitioners to effectively implement STILC by approximating the observed tension fluctuation profile.

The remainder of this paper is organized as follows: Section \ref{sec:2} reviews the physics-based model of the R2R registration error using the perturbation-based approximation method. In Section \ref{sec:3}, the STILC-PID hybrid controller is designed, followed by the convergence analysis. Section \ref{sec:4} demonstrates its performance in terms of registration error control effectiveness and speed of convergence by experimenting with different system parameters in simulations. Section \ref{sec:5} concludes the paper.

\section{R2R Registration Error Dynamics and Disturbaces Modeling}
\label{sec:2}

\subsection{R2R Registration Error Dynamics}
\begin{figure} 
    \centering
    \includegraphics[width=0.48\textwidth]{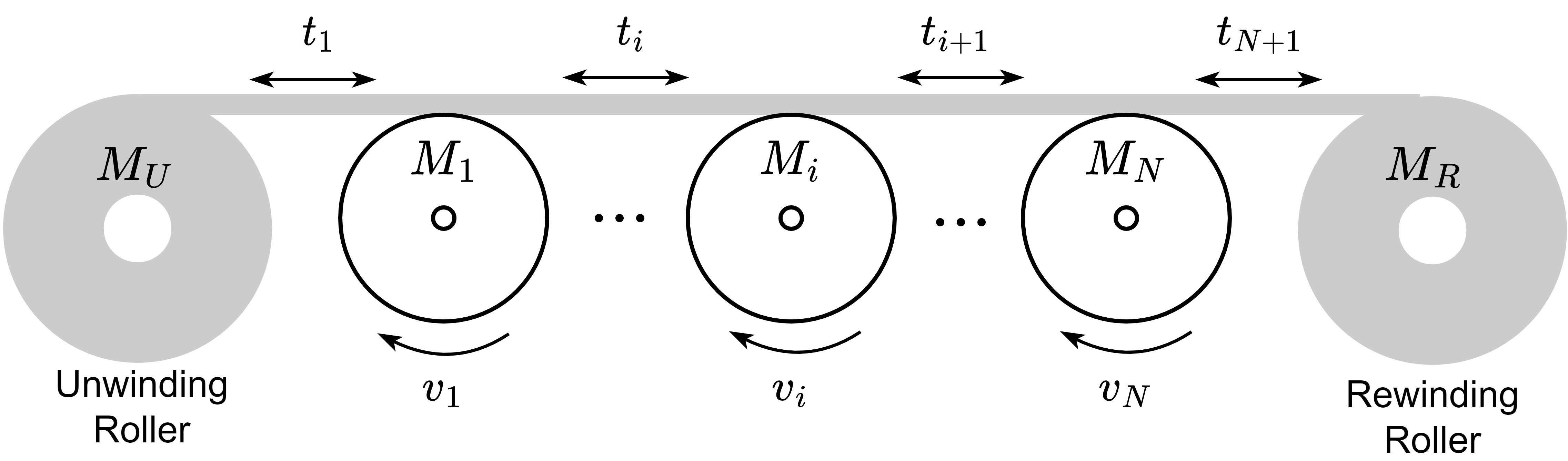}
    \caption{A General R2R Printing System with Unwind, rewind, and Intermediate Rollers}
    \label{fig:R2R System}
\end{figure}

A typical R2R printing system comprises a web handling system that transports the flexible web (substrate) through a series of rollers. In previous work \cite{wang2023spatial,wang2024adaptive}, an R2R printing system with gravure printing rollers was particularly studied, which represents a prevalent configuration of R2R systems in flexible-substrate electronics manufacturing processes. Figure \ref{fig:R2R System} illustrates such an R2R system with one unwinding roller ($M_U$), one rewinding roller ($M_R$), and $N$ intermediate rollers ($M_1$ to $M_N$).

We make the following two assumptions on the system:

\textit{Assumption 1:} There is no slippage between the roller and the web \cite{pagilla2006decentralized,kang2013modeling}. In other words, the tangential roller speeds ($v_i, i=1, 2,\ldots,N$) are equal to the speeds of the following web span. The web span is defined as the web section between two successive rollers. 

\textit{Assumption 2:} The web tension $t_i, i\in\{1,2,\ldots,N+1\}$ and web speed $v_i, i\in\{1,2,\ldots,N\}$ are uniformly distributed through a span.

\begin{figure} 
    \centering
    \includegraphics[width=0.25\textwidth]{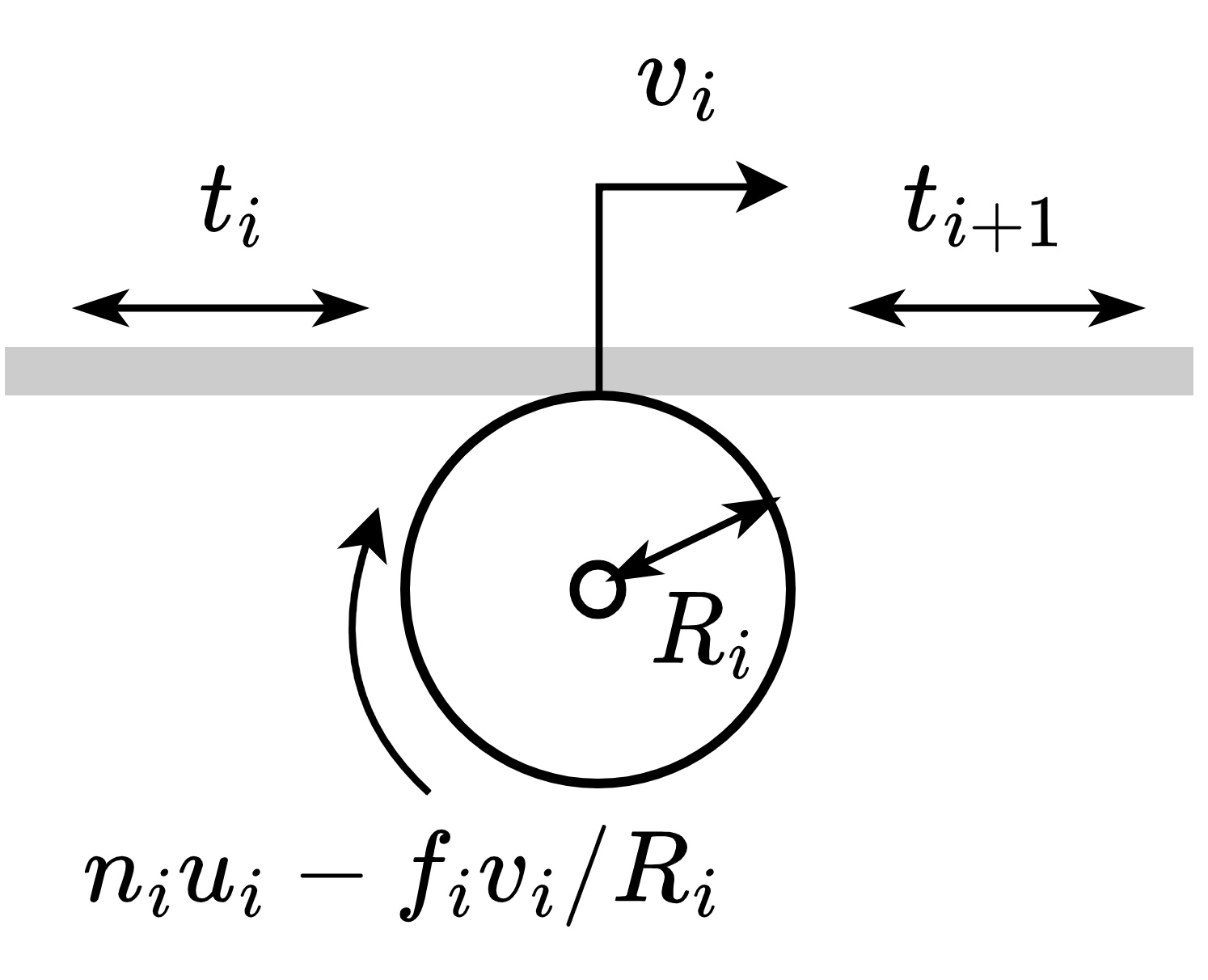}
    \caption{Dynamics of One Intermediate Roller}
    \label{fig:One Roller Dynamics}
\end{figure}

The dynamics of a single motorized intermediate roller in Fig.~\ref{fig:One Roller Dynamics} are described by the following differential equation, considering the mechanics of the rotational system \cite{pagilla2006decentralized}:

\begin{equation}\label{eq:1}
\frac{J_i}{R_i}\dot{v_i}=(t_{i+1}-t_i)R_i+n_iu_i-\frac{f_i}{R_i}v_i
\end{equation}

where $J_i$, $R_i$, $n_i$, and $f_i$  are the inertia, radius, gearing ratio, and friction coefficient of roller $i$, respectively. $u_i$ is the torque input provided by the motor of roller $i$. $i \in \{1,2,\ldots,N\}$ is the index of the roller. It describes how the web behaves according to the controlled motor torque and the inner friction effect.

\begin{figure} 
    \centering
    \includegraphics[width=0.48\textwidth]{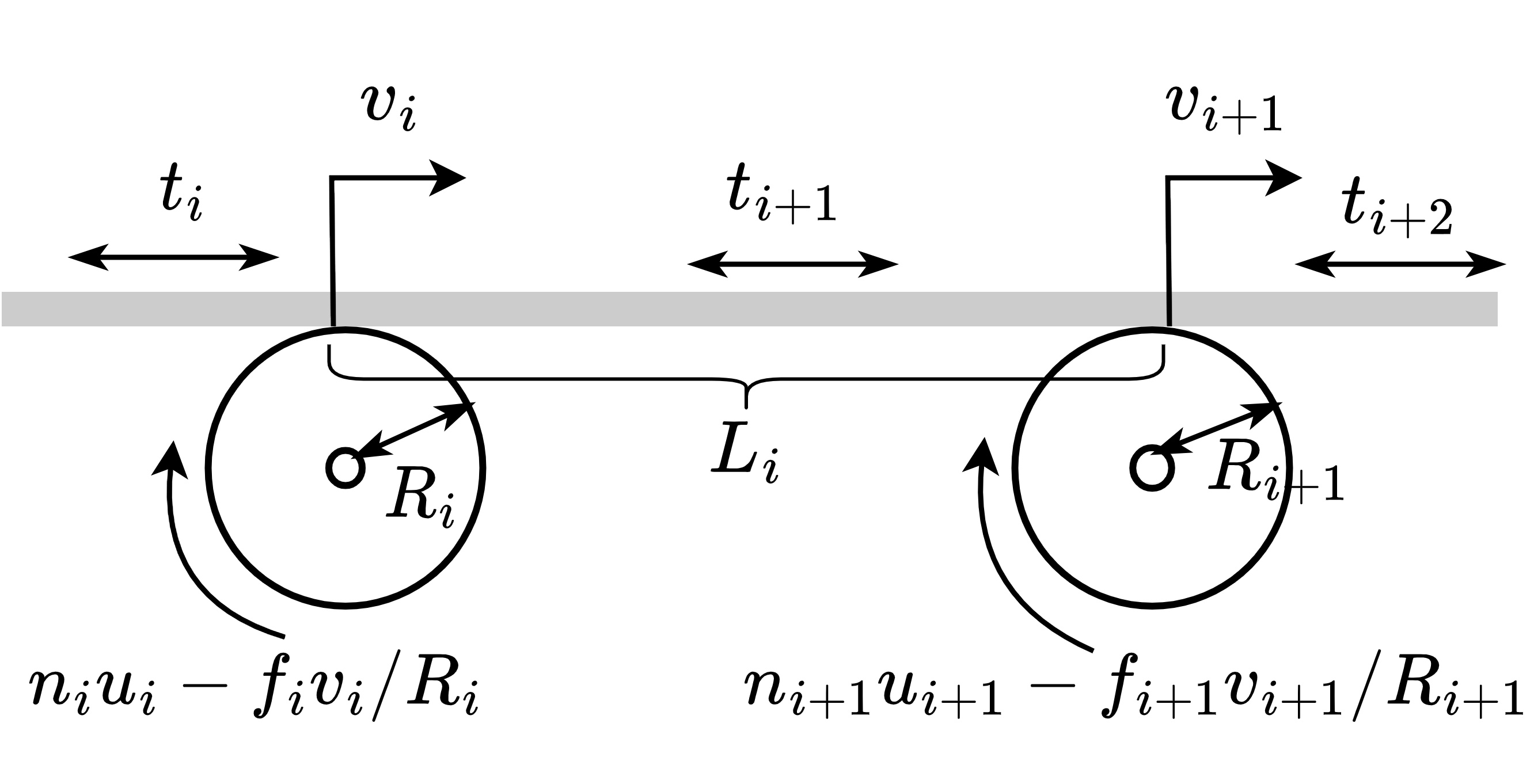}
    \caption{Dynamics of a Two-Roller Printing Unit}
    \label{fig:Two Roller Dynamics}
\end{figure}

Some of the intermediate rollers can be assigned to work as printing rollers. In this paper, the RE is defined as the intended overlay position of a pattern and the actual position printed by two consecutive printing rollers. Thus, we extend the one roller dynamic model in Fig.~\ref{fig:One Roller Dynamics} to the two-roller printing rollers, as shown in Fig.~\ref{fig:Two Roller Dynamics}.

For a two-roller system, we can derive the dynamic equations for $v_i$ and $v_{i+1}$, and the dynamic equations for the tension through the web span between these two rollers and linearize the system with the perturbation method. Details can be found in \cite{pagilla2006decentralized,wang2023spatial,wang2024adaptive}. We just briefly review the perturbation-based system model here.

We define the following perturbation variables:

\begin{equation}\label{eq:3}
\begin{split} 
v_i(\tau)&=v_i^r+V_i(\tau)\\
t_i(\tau)&=t_i^r+T_i(\tau)\\
u_i(\tau)&=u_i^r+U_i(\tau)
\end{split}
\end{equation}

where $\tau$ denotes time. $v_i^r$ and $t_i^r$ are speed and tension references. $V_i$ and $T_i$ are the variations (perturbation variables) in speed and tension. $u_i^r$ is the equilibrium control input to maintain the speed and tension at the given reference levels. $u_i^r$ can be calculated as follows:

\begin{equation}\label{eq:4}
u_i^r=\frac{f_i}{n_iR_i}v_i^r-\frac{R_i}{n_i}(t_{i+1}^r-t_i^r)
\end{equation}

Then we can obtain the dynamic equations for the two-roller case in perturbation form:

\begin{equation}\label{eq:5}
\begin{cases}
    \frac{J_j}{R_j}\dot{V_j}=(T_{j+1}-T_j)R_j+n_jU_j-\frac{f_j}{R_j}V_j \\
    L_k\dot{T_k}=AE(V_k-V_{k-1})\\
                 +(t_{k-1}^rV_{k-1}+v_{k-1}^rT_{k-1})-(t_k^rV_k+v_k^rT_k)\\
                 +(t_{k-1}^rv_{k-1}^r-t_k^rv_k^r) 
\end{cases}
\end{equation}

where $j=i,i+1,$ and $k=i,i+1,i+2$.

Note that Eq.~\eqref{eq:5} includes state variables $v_{i-1}$ and $t_{i-1}$ that are not shown in Fig.~\ref{fig:Two Roller Dynamics} but exist in Fig.~\ref{fig:R2R System}. To simplify the control problem formulation, we assume that in an R2R system shown in Fig.~\ref{fig:R2R System}, tensions and speeds of the rollers except for the printing rollers $M_i$ and $M_{i+1}$ are well-controlled and always equal to the reference values. Therefore, the whole system can be described by Eq.~\eqref{eq:5} with the five state variables ($V_i$, $V_{i+1}$, $T_i$, $T_{i+1}$, $T_{i+2}$). Therefore, in modeling general R2R systems, as shown in Fig.~\ref{fig:R2R System}, we focus exclusively on the two-roller section illustrated in Fig.~\ref{fig:Two Roller Dynamics}.

\begin{figure} 
    \centering
    \includegraphics[width=0.48\textwidth]{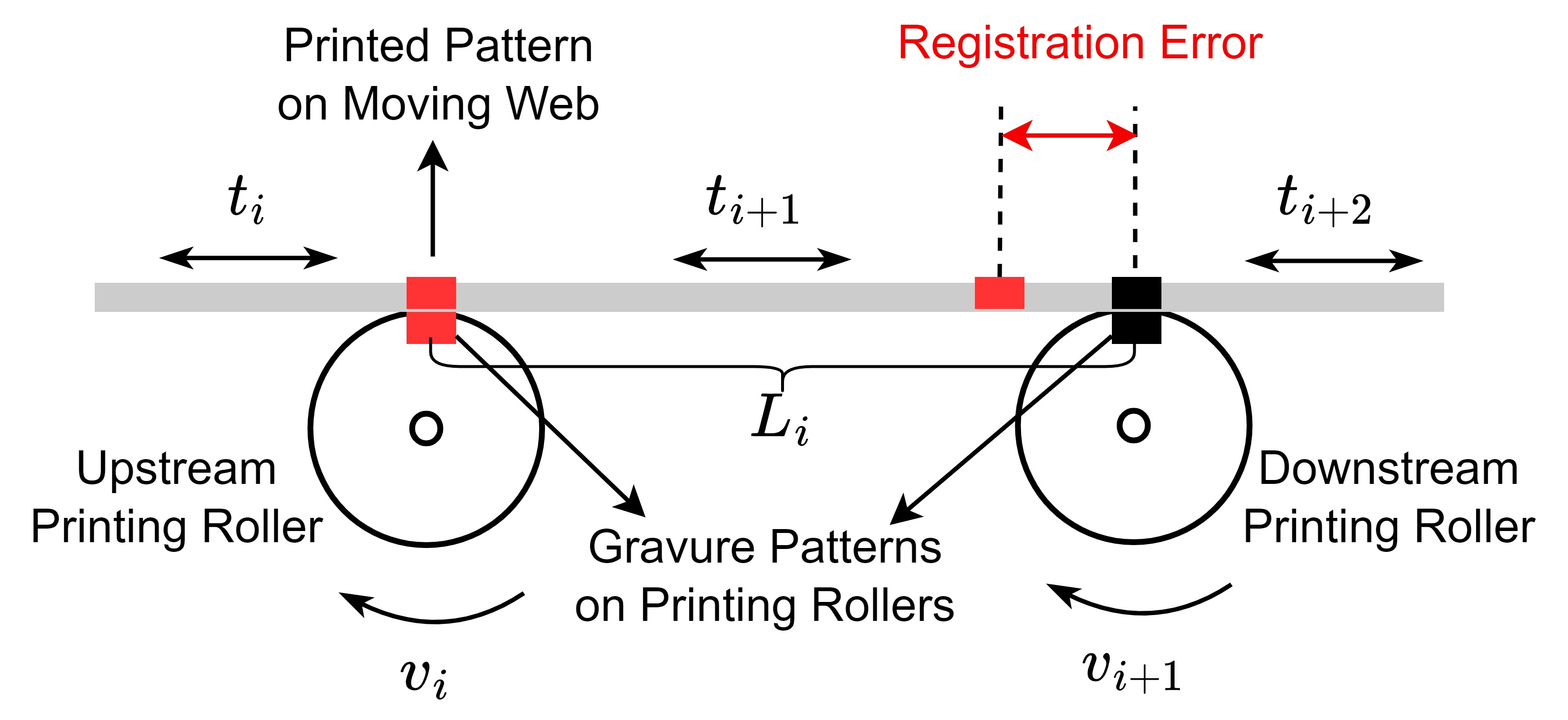}
    \caption{Registration Error in A Two-Roller Printing Unit (The compression rollers are hidden)}
    \label{fig:Registration error}
\end{figure}

After obtaining the state space equations in Eq.~\eqref{eq:5} describing tension and speed variations in the two-roller system, we now designate the two rollers as two adjacent gravure printing rollers and present the dynamic model of the R2R registration error. Figure \ref{fig:Registration error} shows the schematic of two adjacent gravure printing rollers. The red squares represent the patterns printed by the upstream roller. The black square represents the pattern printed by the downstream roller. The registration error is defined as the distance between the two patterns printed by the upstream roller and the downstream roller respectively. Without loss of generality, we make the following assumption on the span length.

\textit{Assumption 3:} The span length between the two printing rollers is equal to the circumference of the upstream roller. ($L_i=2\pi R_i$)

By Assumption 3, we know that when the substrate is running with steady tension and speed, the pattern printed by the downstream roller (black square) should coincide with the pattern printed previously by the upstream roller (red square). Fluctuations in tension and speed due to internal or external disturbances can cause the misalignment between consecutive printed patterns, resulting in registration errors.

Based on \cite{kang2013modeling}, the registration error is given as the following differential equation, which describes the linearized dynamic model of the RE changing rate and the tension and speed variations:

\begin{equation}\label{eq:6}
\dot{r_i}(\tau)=\frac{1}{AE}[v_i^rT_i(\tau-\tau_j^r)-v_{i+1}^rT_{i+1}(\tau)]
\end{equation}

where $r_i$ is the registration error generated by printing roller $i$ and printing roller $i+1$, $\tau_j^r$ is the reference time interval for the upstream printed pattern to be transported to the downstream roller in the $j^{th}$ iteration. Since the variations of tension and speed are relatively small, the actual time interval in each iteration should be very close to a constant value $\tau^r$. Thus, $\tau^r$ is used as the approximated time interval in \cite{kang2013modeling}. In the next section, we will make use of this approximation to rewrite Eq.~\eqref{eq:6} in the iteration domain.

Under Assumptions 1-3, Equations \eqref{eq:5} and \eqref{eq:6} describe the state-space model of the R2R registration error problem as a linear system. The input variables $U_i$ and $U_{i+1}$ are the motor torque variations of the two printing rollers. The output variable is the registration error. The state variables are the variations of the speeds ($V_i, V_{i+1}$) and tensions ($T_i, T_{i+1}$).

\textit{Remark 1:} The process of generating the registration error is repetitive. Each time the pattern is printed, the last operation cycle is terminated and a new operation cycle starts. This is why ILC is considered a suitable approach for such repetitive processes.

\textit{Remark 2:} It should be noted that even though $r_i$ is the variable representing registration error, the actual registration error is only generated every time the downstream pattern is printed. Therefore, the registration error is only measured every time the printing roller completes a cycle. During the cycle, there is no continuous measurement of the registration error. This is the reason for us to design the terminal ILC method for R2R registration control.

\subsection{Modeling of Cycle-Induced Repetitive Disturbances}

From Eq.~\eqref{eq:6}, we know $r(\tau) \equiv 0$ if $T(\tau) \equiv 0$ and $V(\tau) \equiv 0$. However, inherent and exterior disturbances make it difficult to always maintain steady tensions and speeds in an R2R system. In this paper, an axis mismatch is introduced to the upstream roller in the two-roller system in Fig.~\ref{fig:Registration error}. Axis mismatch phenomena commonly occur between the motor shaft and the geometric center of the roller, which causes an angle-periodic disturbance for the repetitive rotary process \cite{xu2008spatial}.

\begin{figure} 
    \centering
    \includegraphics[width=0.48\textwidth]{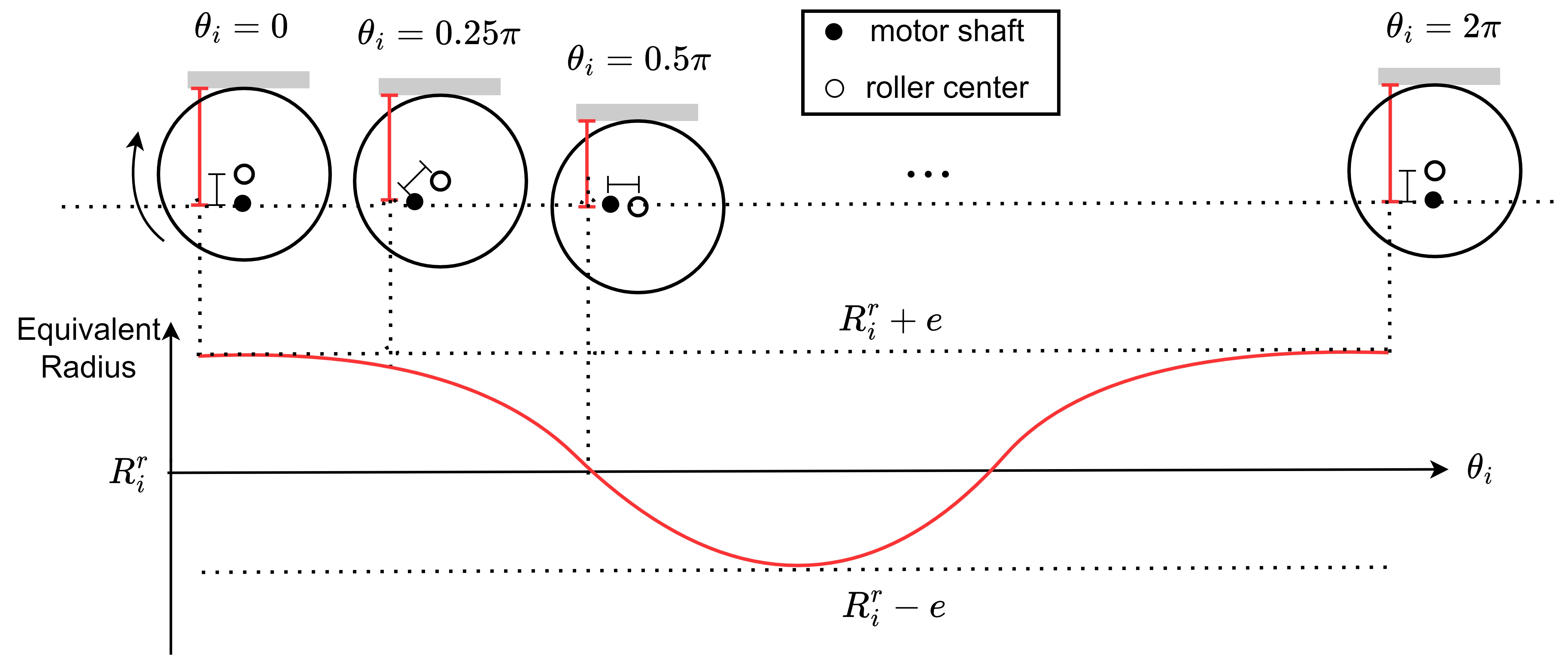}
    \caption{Phase-Angle-Varying Equivalent Radius Caused by Axis Mismatch}
    \label{fig:Varying R}
\end{figure}

When the mismatch exists, the effect of this eccentric printing roller is equivalent to a roller with its radius varying over the phase angle. We define the angle-varying radius as the equivalent radius of the roller with respect to the phase angle. Therefore, the parameter $R_i$ is transferred to a function of the phase angle:

\begin{equation}\label{eq:varying_R}
R_i(\theta_i)=R_i^r+e~cos(\theta_i)
\end{equation}

where $R_i^r$ is the constant value of the original radius when there is no axis mismatch, $e$ is the eccentricity defined as the distance between the motor shaft and the roller center. Figure \ref{fig:Varying R} shows how the equivalent radius $R_i$ varies with respect to the phase angle through an operation cycle.

The axis mismatch introduced to the upstream roller will result in two changes in the previous two-roller system:
(1) The linear time-invariant (LTI) system becomes a linear time-varying (LTV) system due to the varying parameter $R_i$.
(2) The equilibrium control input $u_i^r$ is no longer a constant value but an angle-varying function as shown in Eq.~\eqref{eq:varing_u}.

\begin{equation}\label{eq:varing_u}
u_i^r=\frac{f_i}{n_iR_i(\theta_i(\tau))}v_i^r-\frac{R_i(\theta_i(\tau))}{n_i}(t_{i+1}^r-t_i^r)
\end{equation}

\textit{Assumption 4:} Web tensions and speeds are detectable in real-time, and the registration error of each printing cycle can only be detected at the end of each cycle.

Under Assumptions 1-4, the goal of the following controller design is to minimize the registration error by adjusting the motor torque inputs $u_i$ and $u_{i+1}$. 

\section{Controller Design and Analysis}
\label{sec:3}

\subsection{STILC-PID Hybrid Controller Design}
The controlled system in Sec. \ref{sec:2} is an LTV system with a varying time delay, which makes it difficult to control. We first consider transforming the system into a spatially dependent dynamics. Cobos Torres and Pagilla developed spatially dependent transfer functions for web lateral dynamics that overcome existing limitations by providing web lateral position and slope outputs at any location in the web span in R2R processes \cite{cobos2018spatially}. Hoelzle and Barton introduced a novel framework of analyzing the pixel-dependent dynamics for micro-additive-manufacturing processes using 2-D convolution in spatial coordinates \cite{hoelzle2015spatial}. Practitioners are encouraged to check the details if they have to build spatially dependent dynamics for complex processes.

In this work, we discretize the original system into a discrete-time LTV system with time delay by Zero-Order Holder (ZOH) method. Due to the assumption that the variations of speeds are relatively small in the R2R system described in Sec \ref{sec:2}, the upstream roller rotates by an approximately constant angle, i.e.

\begin{equation}\label{eq:delta_theta}
\delta \theta = \int_{\tau}^{\tau+\delta\tau}v_i(\tau)d\tau \approx v_i^r \delta\tau
\end{equation}

where $\delta\tau$ is the discrete time step. When $\delta\tau$ is relatively small, we can transform the discrete-time system into an approximate discrete-angle system in the spatial domain in terms of $\delta\tau$.

By transforming the original system into a discrete system based on discretized angle step $\delta\tau$, the repetitiveness can be found in the spatial domain and it makes the control problem suitable for ILC methods. Consider in general the angle-varying linear discrete system

\begin{equation}\label{eq:dt_sys}
\begin{split} 
&x_j(\theta+1)=A_j(\theta)x_j(\theta)+B_j(\theta)u_j(\theta) \\
&y_j(\theta)=C_{2,j}(\theta)x_j(\theta)+C_{1,j-1}(\theta)x_{j-1}(\theta)
\end{split}
\end{equation}

 where $\theta=0,1,...,N$, $\theta$ denotes the number of the discretized angle steps and $N$ denotes the total number of the angle steps for a full rotational cycle (i.e. an iteration), and the subscript $j$ indicates the system iteration number. The output $y_j$ is decided by both the states in the current iteration $j$ and the last iteration $j-1$, which corresponds to the time-delay effect in Eq.~\eqref{eq:6}. Thus, the time-delay effect can be converted to the iteration propagation in Eq.~\eqref{eq:dt_sys}.

To simplify the problem, we can assume the system matrices $A_j$, $B_j$, $C_{2,j}$, and $C_{1,j-1}$ are all iteration-invariant when a properly-designed PID controller is applied to the system \eqref{eq:6} to stabilize it. Thus, the stabilized system becomes

\begin{equation}\label{eq:dt_stable_sys}
\begin{split} 
&x_j(\theta+1)=A(\theta)x_j(\theta)+B(\theta)u_j(\theta) \\
&y_j(\theta)=C_2(\theta)x_j(\theta)+C_1(\theta)x_{j-1}(\theta)
\end{split}
\end{equation}

\textit{Remark 3:} TILC methods often require the rigorous repetitiveness of the controlled processes with bounded initial state shifts. In practice, TILC methods require resetting a process to identical initial states before each iteration. This requirement restricts TILC applications to batch processes that can be stopped and restarted, excluding continuous processes that must run uninterrupted. In our work, we use PID to stabilize the system instead of artificial resetting. The repetitiveness requirement and the initial state shift constraint for TILC can be satisfied by the PID component. A TILC component is then added to the control input signal to create a PID-TILC hybrid controller.

The terminal output we want to minimize is the integral of $y_j$ over a fixed angle interval $[0,2\pi]$. When the number of discretization steps $N$ is relatively large, the integral $Y_j$ can be approximated as the following summation:

\begin{equation}\label{eq:Y_j_sum}
\begin{split} 
Y_j&=\sum_{k=1}^N y_j(k)\delta \theta \\
&=\sum_{k=1}^N C_2(k)x_j(k)\delta \theta + \sum_{k=1}^N C_1(k)x_{j-1}(k)\delta \theta
\end{split}
\end{equation}

where $\delta \theta=2\pi/N$. It is reasonable to assume that the $Y_j$ is bounded because the states such as tension and speed variation should be bounded in real-world R2R systems stabilized by properly-tuned PID controllers.

Assume that there exists an equilibrium control input profile $u_{eq}(\theta)$ for a full rotational cycle that can guarantee $x_i(\theta)=0$ for $\theta=0,1,...,N$. In practical applications, we can only know a nominal equilibrium control input $u_{eq}^\ast$ that is a constant vector solved from Eq.~\eqref{eq:5}. Therefore, the introduced axis mismatch disturbance results in an iteration-invariant disturbance profile $u_{dist}(\theta), \theta=0,1,...,N$ to the control input in each iteration, i.e.

\begin{equation}\label{eq:u_dist}
u_{dist}(\theta)=u_{eq}^\ast-u_{eq}(\theta)
\end{equation}

This iteration-invariant disturbance profile $u_{dist}$ will cause a fixed terminal output $Y_{dist}$. When the desired terminal output is set as $Y_d$, the terminal output error $E_j=Y_j-Y_d$ will be a constant value. In the R2R registration error problem, we set $Y_d=0$. Thus, the control goal can be described by the following equation:

\begin{equation}\label{eq:lim_Y_j=0}
\lim_{j\to\infty}Y_j=0
\end{equation}

\textit{Remark 4:} Theoretically, $Y_{dist}$ can be adjusted by tuning the PID controller, but it requires manual tuning work case by case and it is often difficult to realize in practical applications. Also, the PID controller cannot track the change of $Y_d$ when $Y_d$ is not a fixed value through iterations.

To address the problem in Remark 4, we add an STILC component to the control input. The iterative updating law is the P-type ILC updating law with a pre-defined basis function:

\begin{equation}\label{eq:STILC_law}
u_{j+1}^{STILC}(\theta)=u_j^{STILC}(\theta)+\mathcal{L}E_j\Phi(\theta)
\end{equation}

where $u_{j+1}^{STILC}(\theta)$ is the STILC control input component when the phase angle is $\theta$ for the $(j+1)^{th}$ iteration (current iteration), $u_j^{STILC}(\theta)$ is the STILC control input at the same phase angle for the $j$th iteration (last iteration), $\mathcal{L}$ is the learning gain, $E_j$ is the terminal output error generated in the $j^{th}$ iteration (last iteration). In the R2R registration error problem, $E_j$ is the registration error $r_{i,j}$ measured at the end of the $j^{th}$ operation cycle. $\Phi$ is a properly-selected basis function vector

\begin{equation}\label{eq:basis_function}
\begin{split} 
    \Phi(\theta) =& \begin{bmatrix}
    \phi_1(\theta) \\ \phi_2(\theta) \\ \vdots \\ \phi_n(\theta)
    \end{bmatrix}
\end{split}
\end{equation}

where $n$ is the dimension of the input vector, $\phi_1(\theta),\phi_2(\theta),...,\phi_n(\theta)$ are $R\to R$ functions.

\subsection{Stability Analysis}

By solving Eq.~\eqref{eq:dt_stable_sys}, we obtain the transition equation from the initial state

\begin{equation}\label{eq:transition_eq}
x_j(k)=G(k)x_j(0)+H(k)\Xi_j+H_d(k)
\end{equation}

where $G(k)$ and $H(k)$ cab be obtained by the following recursive calculations:

\begin{equation}\label{eq:recursion}
\begin{split} 
&G(k+1)=A(k)G(k), G(0)=I,\\
&H(k+1)=A(k)H(k)+B(k)\Phi(k), H(0)=0,\\
&H_d(k+1)=A(k)H_d(k)+B(k)u_{dist}(k),\\
&H_d(0)=B(0)u_{dist}(0), k=0,1,...,N-1
\end{split}
\end{equation}

and $\Xi_j$ can be obtained by the following iterative updating equation:

\begin{equation}\label{eq:Xi_update}
\Xi_j=\Xi_{j-1}+\mathcal{L}E_{j-1}, \Xi_0=0, j=1,2,...
\end{equation}

Substituting \eqref{eq:transition_eq} into \eqref{eq:Y_j_sum}, we get

\begin{equation}\label{eq:Y_i_detail}
\begin{split} 
Y_j &= [\sum_{k=1}^N C_2(k)G(k)x_j(0) + \sum_{k=1}^N C_1(k)G(k)x_{j-1}(0)]\delta \theta \\
& + \sum_{k=1}^N [C_2(k)+C_1(k)]H_d(k)\delta \theta \\
& + [\sum_{k=1}^N C_2(k)H(k)+\sum_{k=1}^N C_1(k)H(k)]\delta \theta \Xi_{j-1} \\
& + \sum_{k=1}^N C_2(k)H(k)\delta \theta \mathcal{L} E_{j-1}
\end{split}
\end{equation}

If the desired terminal output $Y_d$ is given, based on \eqref{eq:transition_eq}\eqref{eq:recursion}\eqref{eq:Xi_update} we can get

\begin{equation}\label{eq:E_eq_1}
\begin{split} 
E_j &= Y_j - Y_d \\
&= \Omega_1 E_{j-1} + \Omega_2 \sum_{s=1}^{j-2} E_s +\Omega_3
\end{split}
\end{equation}

where
\begin{equation}\label{eq:Omega}
\begin{split} 
&\Omega_1=\sum_{k=1}^N C_2(k)H(k)\delta \theta \mathcal{L} \\
&\Omega_2=[\sum_{k=1}^N C_2(k)H(k)+\sum_{k=1}^N C_1(k)H(k)]\delta \theta \\
&\Omega_3=[\sum_{k=1}^N C_2(k)G(k)x_j(0) + \sum_{k=1}^N C_1(k)G(k)x_{j-1}(0)]\delta \theta + \\
&\sum_{k=1}^N [C_2(k)+C_1(k)]H_d(k)\delta \theta -Y_d
\end{split}
\end{equation}

In the following theorem, we show that $E_j$ converges to zero when $\Omega_1$ and $\Omega_2$ satisfy a specific criterion. This theorem represents a pivotal finding in our research, as it establishes the fundamental conditions under which R2R systems can effectively reduce registration errors. This theorem applies to multi-layer manufacturing processes that are governed by the LTV system \eqref{eq:dt_stable_sys} and the second-order terminal output dynamics \eqref{eq:Y_j_sum}. Different from typical ILC problems where resetting operation is applied to restoring a constant initial state for each iteration, the continuously running R2R system cannot clean up the information generated in the last iteration by resetting at the end of the operation cycle. The information in the last iteration (e.g. the printed position or the deposition height of the last layer) is propagated to the current iteration and influences the terminal output for the current iteration. By applying STILC to these processes, we derive the convergence criterion by analyzing the convergence of a second-order recurrence series, as described in \eqref{eq:E_eq_recurrence}.

\textit{Theorem 1: Consider a stabilized discrete LTV system \eqref{eq:dt_stable_sys} and a given achievable terminal output $Y_d$ integrated as \eqref{eq:Y_j_sum}. Through the repetitive operation cycles, the STILC law in \eqref{eq:STILC_law} will make the terminal error $E_j$ converge to zero if $\lvert \lambda_1\rvert <1$ and $\lvert \lambda_2\rvert <1$, where $\lambda_1$, $\lambda_2$ are the two roots of the following characteristic equation: }

\begin{equation}\label{eq:characteristic_eq}
\lambda^2-(\Omega_1+1)\lambda+(\Omega_1-\Omega_2)=0.
\end{equation}

The proof of Theorem 1 is provided in Appendix A.
~\\

\textit{Remark 5:} From Theorem 1, we know the convergence performance of the terminal error $E_j$ is decided by two parameters, $\Omega_1$ and $\Omega_2$. When designing the controller, we can design different basis function matrices $\Phi$ and learning gains $\mathcal{L}$ to tune $\Omega_1$ and $\Omega_2$. Based on Theorem 1, we can derive some specific rules to help us design the appropriate basis function vector and learning gain.

\subsection{Basis Function Design and Learning Gain Selection}

Ideally, we expect the designed STILC to fully compensate $u_{dist}$ once it detects any terminal error so that any fluctuations in tensions or speeds can be fully eliminated, as well as the registration error. As we define $\Xi_0=0$ in \eqref{eq:Xi_update}, the STILC will not generate any compensation signal in the first iteration. At the end of the first iteration, it detects the terminal error and starts to generate a compensation signal for the second iteration. Thus, the ideal $\Phi$ and $\mathcal{L}$ should satisfy the following relation:

\begin{equation}\label{eq:ideal_Phi_L}
\Phi(k)\mathcal{L}E_1=-u_{dist}(k), k=0,1,...,N-1.
\end{equation}

Thereafter, any fluctuations or the terminal error can be eliminated after just one iteration. However, it is difficult to obtain an exact $u_{dist}$ profile in practical applications. In the following, we state that it is acceptable if an approximate basis function vector can be designed based on some prior knowledge or observations about the process. For example, in a rotary system with axis mismatches, we can use trigonometric functions as the basis functions.

The transition equation \eqref{eq:transition_eq} can be rewritten to include $u_j$ and $u_{dist}$ explicitly:

\begin{equation}\label{eq:tilde_u}
x_j(k) = G(k)x_j(0) + \mathcal{H}(k) [\Tilde{u}_j(k) + \Tilde{u}_{dist}]
\end{equation}

where $\Tilde{u}_j(k)=[u_j(0)\  u_j(1)\  u_j(2)\  \cdots\  u_j(k-1)\  0\  \cdots]^T$ is a truncated profile of control input $u$. $\Tilde{u}_{dist}$ is similar. And we know

\begin{equation}\label{eq:tilde_u_tilde_Phi}
\Tilde{u}_j(k) = \Tilde{u}_0(k) + \Xi_j\Tilde{\Phi}(k)
\end{equation}

where $\Tilde{\Phi}(k)=[\Phi(0)\  \Phi(1)\  \Phi(2)\  \cdots\  \Phi(k-1)\   0\  \cdots]^T$.

We can also write $\Tilde{u}_j(k)$, $\Tilde{u}_{dist}$, and $\Tilde{\Phi}(k)$ as the multiplication of a matrix $M_k$ and the complete profiles:

\begin{equation}\label{eq:u_phi_tilde_matrix_form}
\begin{split} 
&\Tilde{u}_j(k) = M_k u_j \\
&\Tilde{u}_{dist} = M_k u_{dist} \\
&\Tilde{\Phi}(k) = M_k \Phi 
\end{split}
\end{equation}

where $M_k$ is

\begin{equation}\label{eq:M_k}
M_k = \begin{bmatrix}
    I_k & \mathbf{0} \\
   \mathbf{0} & \mathbf{0}
\end{bmatrix}_{N \times N}
\end{equation}

Rearranging \eqref{eq:Y_i_detail}, we can obtain

\begin{equation}\label{eq:Y_i_rearrange}
\begin{split} 
Y_j/\delta \theta &= \sum_{k=1}^N[C_2(k)G(k)x_j(0) + C_1(k)G(k)x_{j-1}(0)] \\
& + \sum_{k=1}^N[C_2(k)\mathcal{H}(k)M_k]\Delta u_{j-1} \\
& + \sum_{k=1}^N[C_2(k)\mathcal{H}(k)M_k+C_1(k)\mathcal{H}(k)M_k](\Xi_{j-1}\Phi+u_{dist})
\end{split}
\end{equation}

Note that $\Xi$ is a scalar and $\Phi$ is a $N \times 1$ vector in this work. The STILC updating law aims to approach a $\Xi$ iteratively making $Y$ converge to the desired value (zero in this problem). Let $\mathcal{N}=\sum_{k=1}^N[C_2(k)\mathcal{H}(k)M_k+C_1(k)\mathcal{H}(k)M_k]$. We can confirm the existence of such a $\Xi$ as stated in the following theorem:

~\\

\textit{Theorem 2: If the designed basis function vector $\Phi$ satisfies $\mathcal{N}\Phi \neq 0$, then there exists a $\Xi$ such that $Y = 0$.}

This conclusion is obvious because this is a question about the existence of solutions for a first-degree linear equation with one variable.

In this problem, we can further assume that the initial states of iterations, $x_{j-1}(0)$ and $x_j(0)$, are zero. When convergence has been achieved, the second term in \eqref{eq:Y_i_rearrange} is also zero because $\Delta u_{j-1}=0$. If $\Phi$ and $u_{dist}$ are parallel vectors, it is evident that there exists a solution $\Xi$ making $(\Xi\Phi+u_{dist})$ zero. If there is a small angle between $\Phi$ and $u_{dist}$, we should still be able to find a solution $\Xi$ making $(\Xi\Phi+u_{dist})$ orthogonal to $\mathcal{N}$. The process of convergence in the iteration domain becomes solving a linear equation with one unknown variable, which is considered easy to solve numerically. This is the fundamental reason for the simplicity of the proposed STILC design.

Compared with designing an ideal basis function vector $\Phi$ in \eqref{eq:ideal_Phi_L} that generates a completely opposite signal to $u_{dist}$, it is easier for practitioners to tune the learning gain $\mathcal{L}$ to make $\Omega_1$ and $\Omega_2$ satisfy the criteria of convergence in Theorem 1.

Solving \eqref{eq:characteristic_eq}, we can get the two characteristic roots:

\begin{equation}\label{eq:characteristic_roots}
\begin{split} 
&\lambda_1=\frac{1}{2}[\Omega_1+1+\sqrt{(\Omega_1-1)^2+4\Omega_2}] \\
&\lambda_2=\frac{1}{2}[\Omega_1+1-\sqrt{(\Omega_1-1)^2+4\Omega_2}]
\end{split}
\end{equation}

For the two roots, there are the following two cases.
~\\

\noindent\textit{Case 1: $(\Omega_1-1)^2+4\Omega_2 \geq 0$}\par
In this case, $\lambda_1$ and $\lambda_2$ are real numbers. In order to satisfy $\lvert \lambda_1\rvert <1$ and $\lvert \lambda_2\rvert <1$, we need

\begin{equation}\label{eq:characteristic_roots_case_1}
\begin{split} 
&\frac{1}{2}[\Omega_1+1+\sqrt{(\Omega_1-1)^2+4\Omega_2}]<1 \\
&\frac{1}{2}[\Omega_1+1-\sqrt{(\Omega_1-1)^2+4\Omega_2}]>1.
\end{split}
\end{equation}

From \eqref{eq:characteristic_roots_case_1} and the condition for Case 1, we obtain $-3<\Omega_1<1$ and $\Omega_2<0$.
~\\

\noindent\textit{Case 2: $(\Omega_1-1)^2+4\Omega_2 < 0$}\par
In this case, $\lambda_1$ and $\lambda_2$ are complex numbers. In order to satisfy $\lvert \lambda_1\rvert <1$ and $\lvert \lambda_2\rvert <1$, we need

\begin{equation}\label{eq:characteristic_roots_case_2}
\frac{1}{4}(\Omega_1+1)^2-\frac{1}{4}[(\Omega_1-1)^2+4\Omega_2]<1 
\end{equation}

From \eqref{eq:characteristic_roots_case_2} and the condition for Case 2, we obtain $\Omega_1<1$ and $\Omega_2<0$.
~\\

It is evident that $\Omega_2<0$ is required in both cases. Therefore, once an approximate basis function vector $\Phi$ is defined, the first critical step in designing the learning gain $\mathcal{L}$ is to determine the correct sign for $\mathcal{L}$. An incorrect sign will prevent any possibility of achieving convergence. Then we can increase the absolute value of $\mathcal{L}$ as much as possible, as a higher learning gain enhances the responsiveness of STILC and increases the learning speed. However, setting the learning gain too high can lead to significant overshoot or even result in divergence. To prevent these issues, we set the learning gain to satisfy $(\Omega_1-1)^2+4\Omega_2 = 0$. In section \ref{sec:4}, we show the effects of different learning gain selections.

\section{Simulation Verification}
\label{sec:4}
To thoroughly verify and validate the proposed STILC-PID method, We develop a numerical model of a two-roller R2R printing system in Simulink and configure the parameters as shown in Table \ref{tab:1}. The simulation results are then discussed.

\begin{table}[t]
\caption[Table]{Simulation Parameters\label{tab:1}}
\scalebox{0.85}{
\begin{tabular}{lll}
\hline
Parameter & Notation & Value \\
\hline
Cross-sectional Area & $A$ & $1.29 \times 10^{-5}$ m$^2$\\
Young's Modulus & $E$ & 186.158 MPa \\
Reference Roller Radius & $R_i^r$,$R_{i+1}$ & 0.381 m \\
Inertia of Roller & $J_i$,$J_{i+1}$ & 0.146 kg$\cdot$m$^2$\\
Friction Coefficient & $f_i$,$f_{i+1}$ & 0.685 \\
Gear Ratio & $n_i$,$n_{i+1}$ & 1 \\
Span Length & $L_i$,$L_{i+1}$,$L_{i+2}$ & 2.4 m \\
Reference Speed & $v_i^r$,$v_{i+1}^r$ & 0.16 m/s \\
Reference Tension & $t_i^r$,$t_{i+1}^r$,$t_{i+2}^r$ & 20 N \\
Reference Period Time & $\tau_i^r$ & 14.962 s \\
\hline
\end{tabular}
}
\end{table}

\subsection{Non-Zero RE Convergence Performance from Feedback Control Methods}
\begin{figure} 
    \centering
    \includegraphics[width=0.42\textwidth]{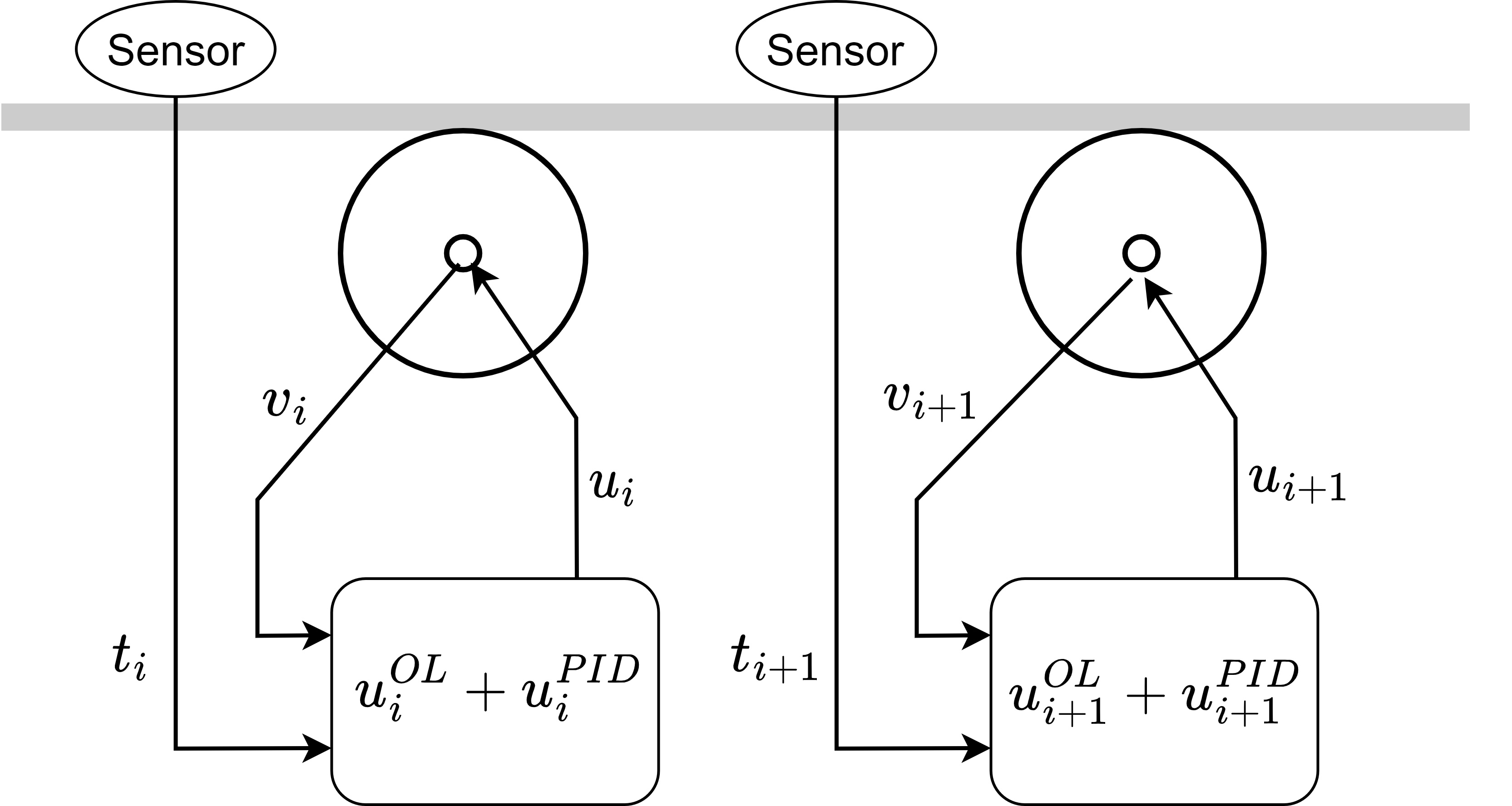}
    \caption{Decentralized PID Control for Tension and Speed Regulation}
    \label{fig:PID}
\end{figure}

In \cite{pagilla2006decentralized}, a decentralized controller is designed to regulate the speeds and tensions in the R2R system. Figure \ref{fig:PID} shows the schematic of the decentralized control scheme for a two-roller system. The control input signal for each roller motor is the summation of two components: (a) the open-loop component given by Eq.~\eqref{eq:4}, and (b) the closed-loop component generated by a PID controller.

The control law of the decentralized controller for gravure printing roller $i$ and roller $(i+1)$ is given as the following:

\begin{equation}\label{eq:D_PID}
\begin{split}
    &u_i(\tau)=u_i^{OL}+u_i^{PID}(\tau)\\
    &u_i^{OL}=u_i^r\\
    &u_i^{PID}(\tau)=K_i^P{\begin{bmatrix}
    T_i(\tau) & V_i(\tau)
    \end{bmatrix}
    }^T \\
    &~~~~~~~~~~~ +K_i^I{\begin{bmatrix}
    \int_0^\tau T_i(\tau) d\tau & \int_0^\tau V_i(\tau) d\tau
    \end{bmatrix}
    }^T \\
    &~~~~~~~~~~~ +K_i^D{\begin{bmatrix}
    \dot{T_i}(\tau) & \dot{V_i}(\tau)
    \end{bmatrix}
    }^T \\
    &u_{i+1}(\tau)=u_{i+1}^{OL}+u_{i+1}^{PID}(\tau)\\
    &u_{i+1}^{OL}=u_{i+1}^r\\
    &u_{i+1}^{PID}(\tau)=K_{i+1}^P{\begin{bmatrix}
    T_{i+1}(\tau) & V_{i+1}(\tau)
    \end{bmatrix}
    }^T \\
    &~~~~~~~~~~~ +K_{i+1}^I{\begin{bmatrix}
    \int_0^\tau T_{i+1}(\tau) d\tau & \int_0^\tau V_{i+1}(\tau) d\tau
    \end{bmatrix}
    }^T \\
    &~~~~~~~~~~~ +K_{i+1}^D{\begin{bmatrix}
    \dot{T}_{i+1}(\tau) & \dot{V}_{i+1}(\tau)
    \end{bmatrix}
    }^T
\end{split}
\end{equation}

where $u_i^{OL}$ and $u_{i+1}^{OL}$ are the open-loop input components, $u_i^{PID}$ and $u_{i+1}^{PID}$ are the closed-loop input components provided by the decentralized PID, and $K_i$ and $K_{i+1}$ are the feedback gain vectors.

\begin{figure}
\centering
\begin{subfigure}[b]{0.42\textwidth}
    \includegraphics[width=\textwidth]{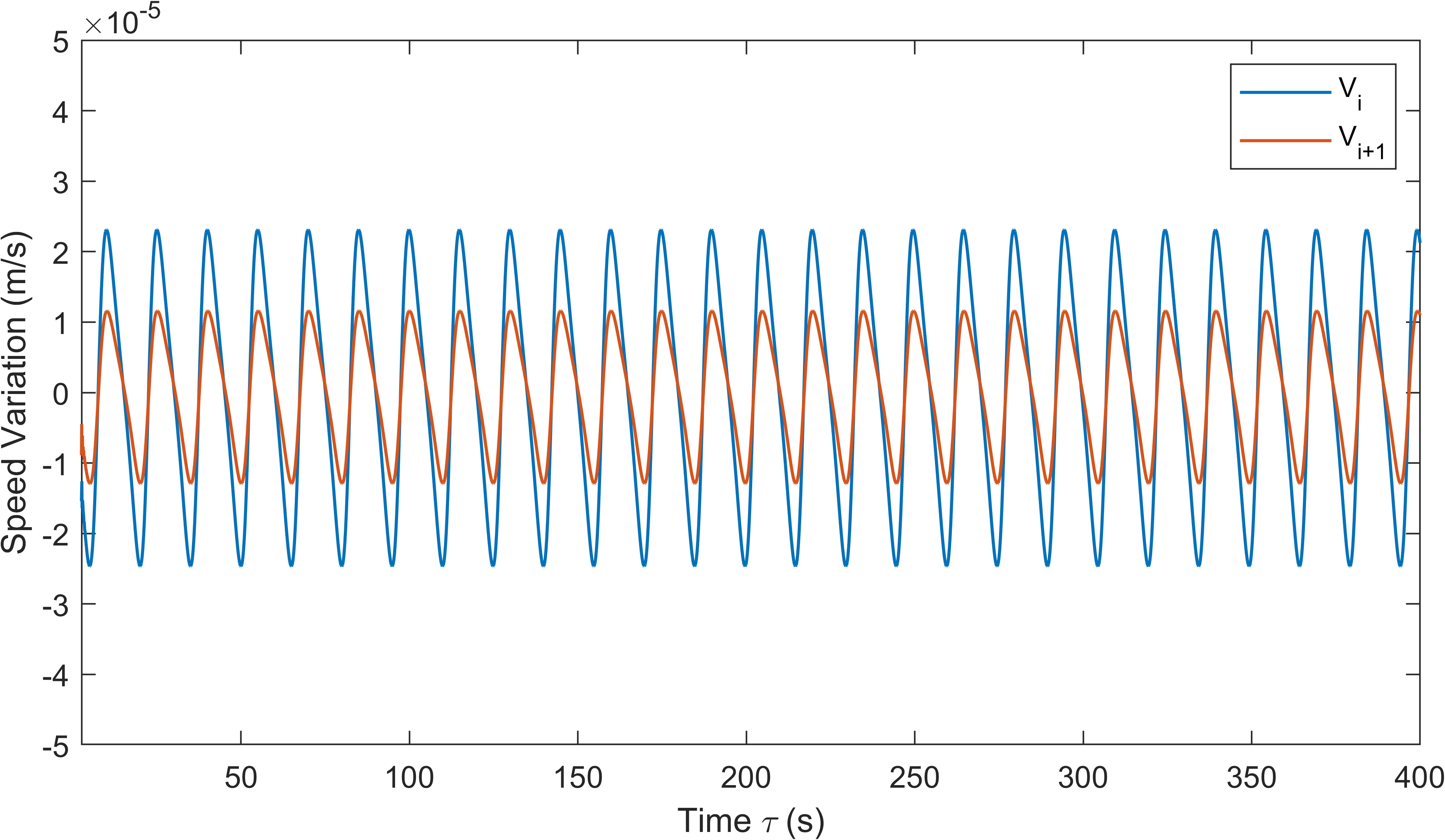}
    \caption{}
    \label{fig:TV_fluctuation_a}
\end{subfigure}
\hfill
\begin{subfigure}[b]{0.42\textwidth}
    \includegraphics[width=\textwidth]{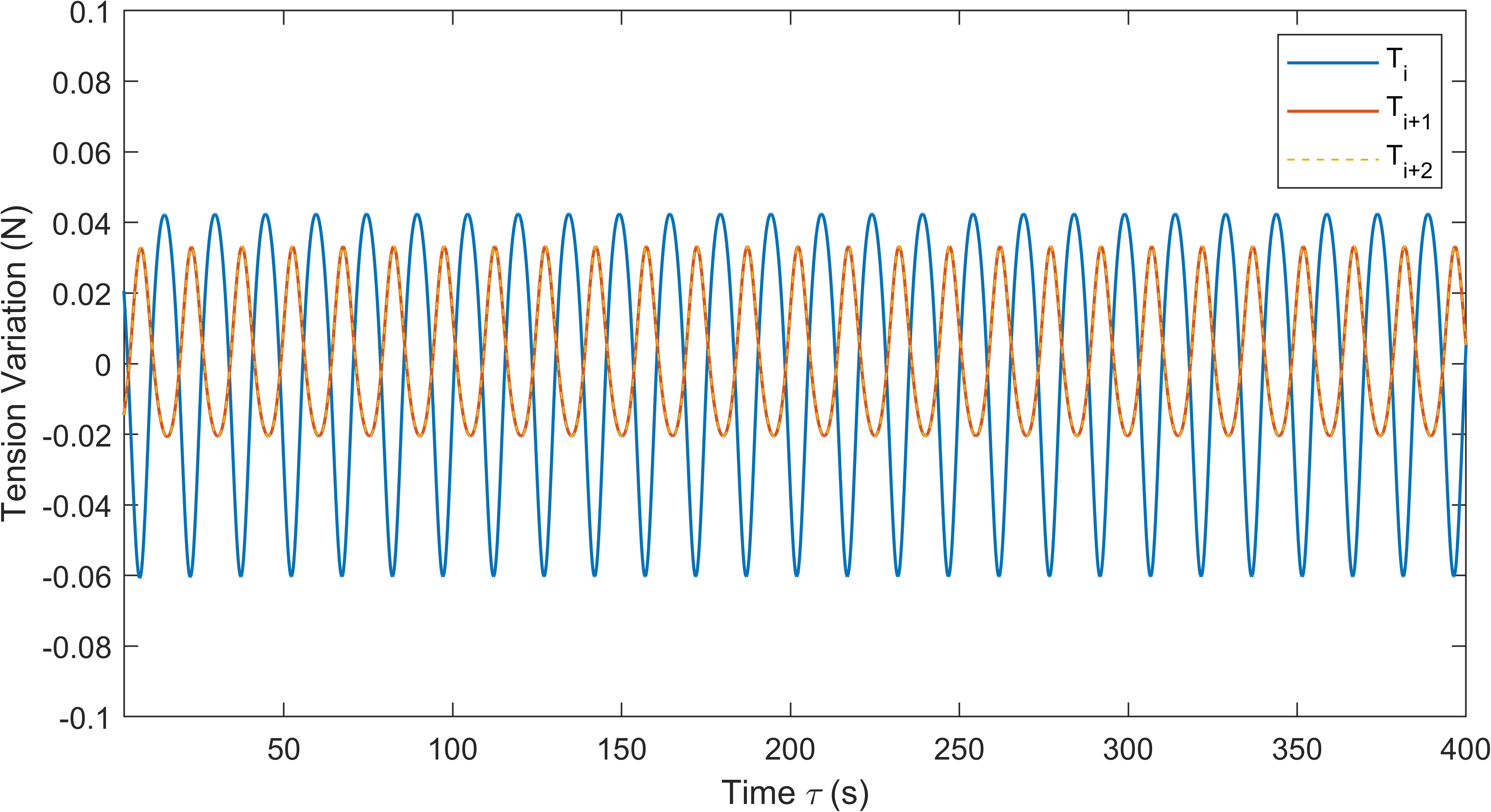}
    \caption{}
    \label{fig:TV_fluctuation_b}
\end{subfigure}
\caption{Speed and Tension Variations Caused by Axis Mismatch. (a)Speed Variations; (b)Tension Variations}
\label{fig:TV_fluctuation}
\end{figure}

\begin{figure}
\centering
\begin{subfigure}[b]{0.45\textwidth}
    \centering
    \includegraphics[width=\textwidth]{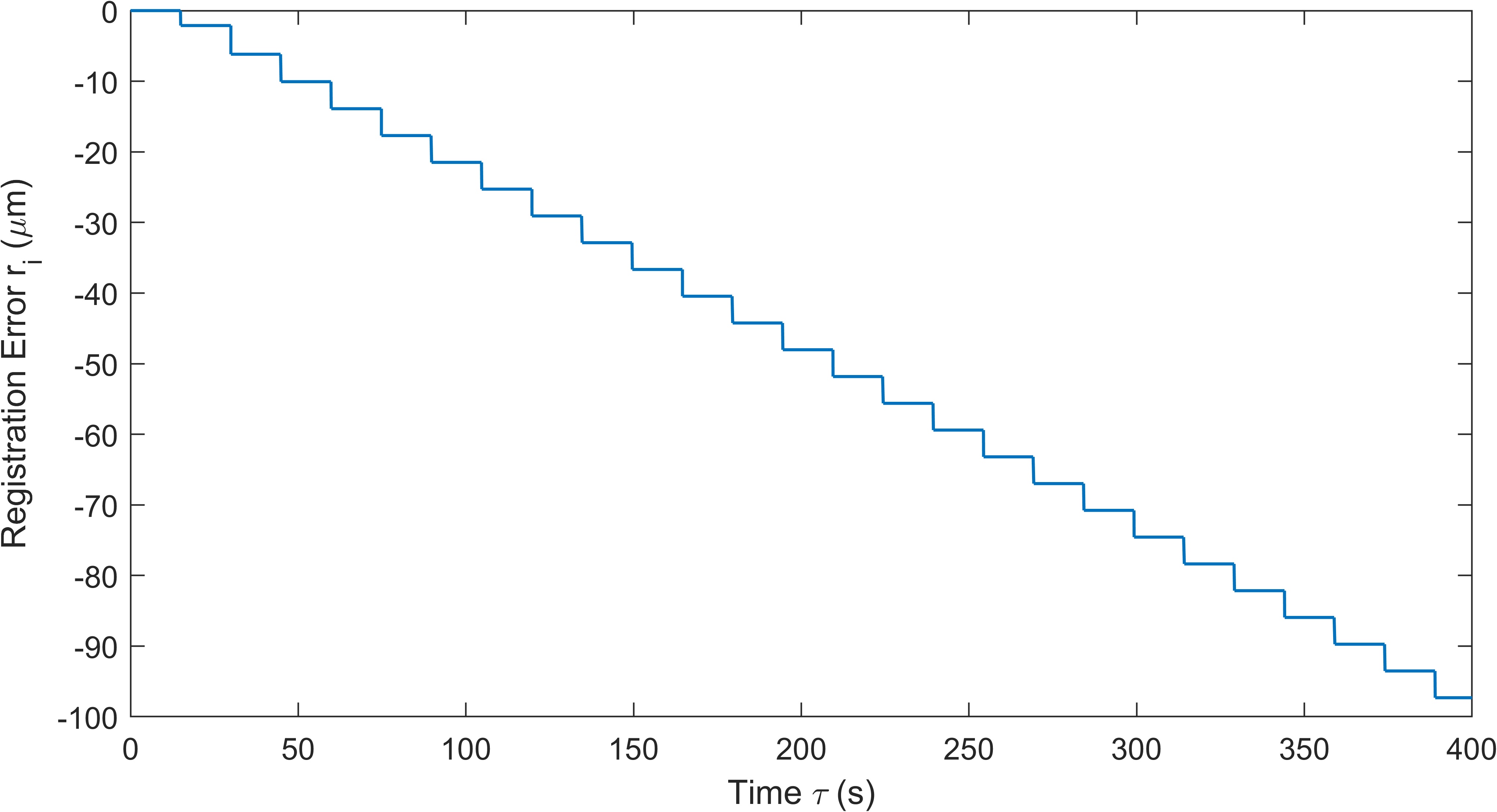}
    \caption{}
    \label{fig:RE_OL_a}
\end{subfigure}
\hfill
\begin{subfigure}[b]{0.45\textwidth}
    \centering
    \includegraphics[width=\textwidth]{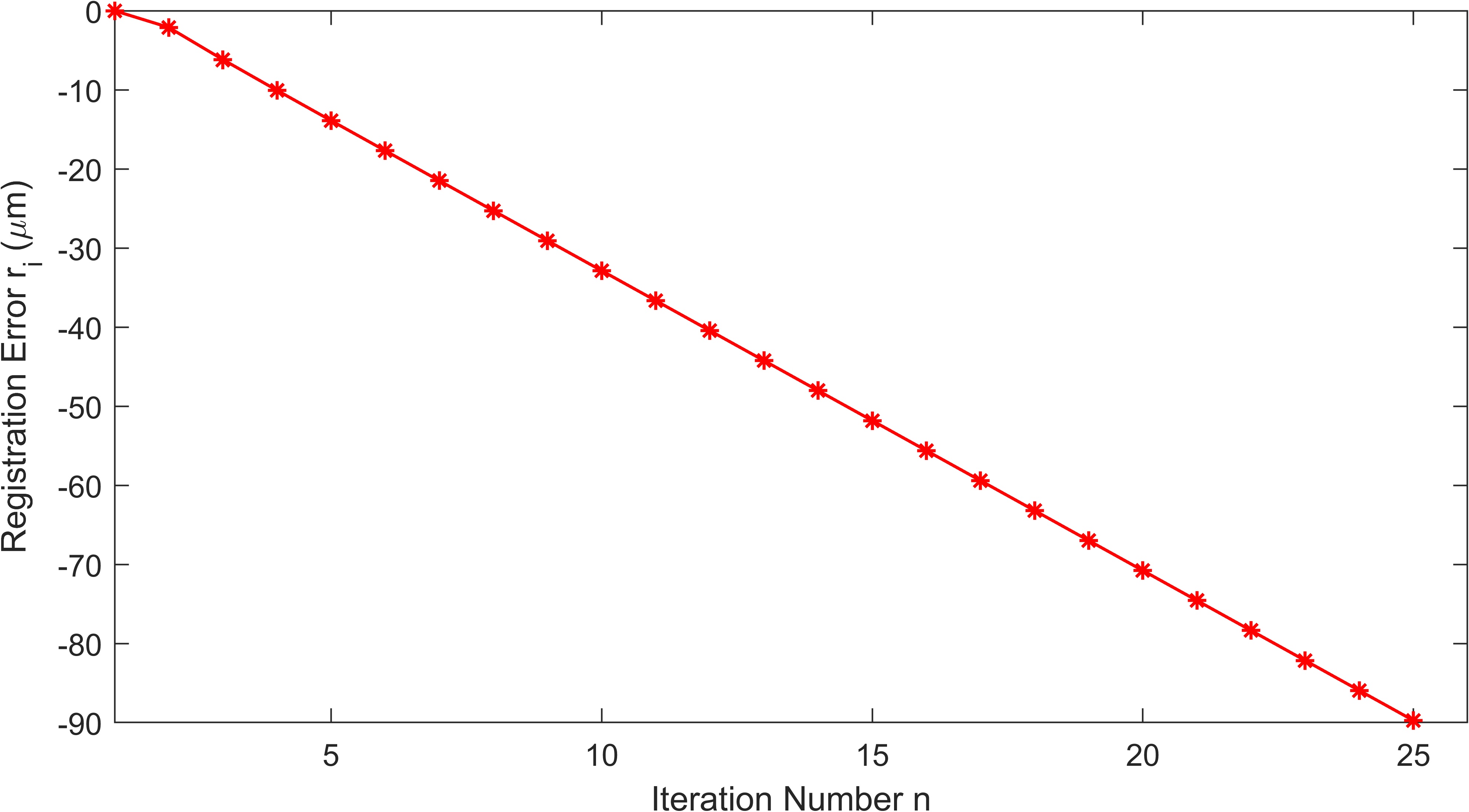}
    \caption{}
    \label{fig:RE_OL_b}
\end{subfigure}
\caption{Registration Error Caused by Axis Mismatch with Only Open-loop Control Input. (a) The actual registration error value in time domain; (b) Registration error in the iteration domain}
\label{fig:RE_OL}
\end{figure}

Figure \ref{fig:TV_fluctuation} illustrates the temporal fluctuations of the five state variables described by Eq.~\eqref{eq:5} when an axis mismatch is introduced to the upstream roller. It should be noted that $T_{i+1}$ and $T_{i+2}$ overlap in this case. On the other hand, Figure \ref{fig:RE_OL} displays the registration error resulting from speed and tension fluctuations. It is important to emphasize that $r_i(\tau)$ represents a continuous function of time; however, it can only be measured by the sensor after the printing of the downstream pattern on the web. Consequently, the actual registration error is updated whenever the terminal condition is met, specifically when the phase angle of the gravure pattern on the downstream roller reaches $2\pi$. Throughout the intermediate process of each iteration, the registration error remains constant after the initial update at the beginning of the iteration. As depicted in Figure \ref{fig:RE_OL_b}, the registration error accumulates and dilates iteratively when only the open-loop input component is applied to the system.

\begin{figure} 
    \centering
    \includegraphics[width=0.45\textwidth]{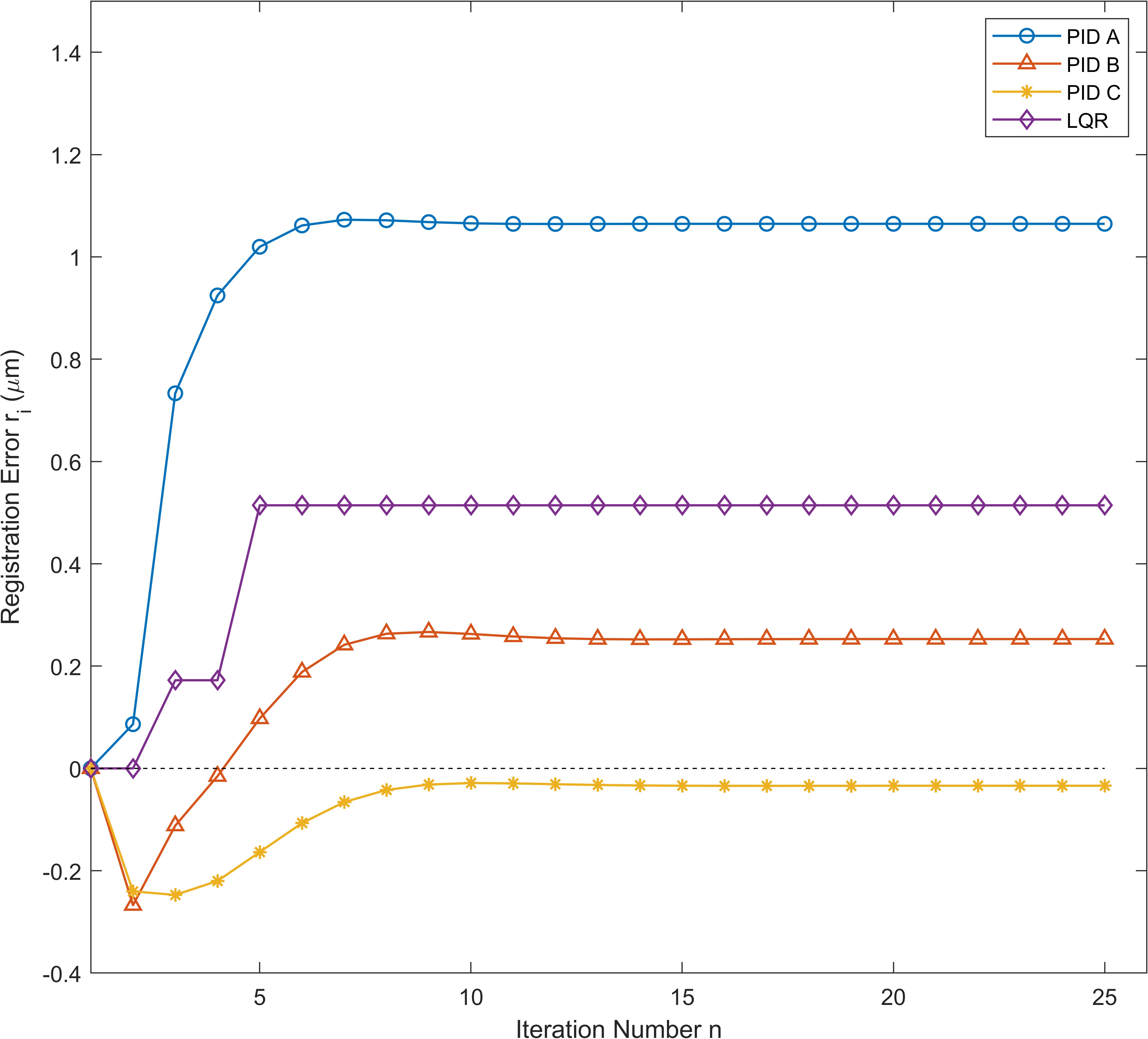}
    \caption{Registration Error Controlled by Decentralized PID Controllers with Different Parameter Settings}
    \label{fig:Compare_PID}
\end{figure}

Then we compare the effects of decentralized PID controllers with different parameter settings. Given the similar dynamics of the upstream and downstream rollers, the two PID modules for the two rollers share the same parameter setting. In Figure \ref{fig:Compare_PID}, we compare the performances of three decentralized PID controllers with different parameter settings. We tuned 3 different PID controllers empirically and their parameters are listed in Table \ref{tab:PID_Settings}. Controller PID A exhibits the highest convergence error, demonstrating the least effective performance. In contrast, PID B and PID C show significantly improved performance, with PID C achieving a particularly low registration error level. Nonetheless, it should be noted that while the registration errors converge to constant values, convergence to zero is not assured. This observation aligns with the analysis presented in Remark 4. In engineering practice, it can be challenging for practitioners to find an optimal set of PID parameters that will precisely converge the registration error to zero. Additionally, the optimal PID parameters may need to be adjusted every time the system parameters change.

\begin{table}[!t]
\caption[Table]{Controller Settings\label{tab:PID_Settings}}
\scalebox{0.85}{
\begin{tabular}{lccc}
\hline
Setting & $K^P$ & $K^I$ & $K^D$ \\
\hline
PID A & $[-0.1916 ~ 0]$ & $[-0.0767 ~ 0]$ & $[-0.0038 ~ -0.1916]$  \\
PID B & $[-0.1916 ~ 0]$ & $[-0.0575 ~ 0]$ & $[-0.0038 ~ -0.1916]$ \\
PID C & $[-0.3832 ~ 0]$ & $[-0.0575 ~ 0]$ & $[-0.0038 ~ -0.1916]$ \\
\hline
\end{tabular}
}
\end{table}

Besides PID control, LQR is also a popular class of feedback control strategy to minimize a quadratic-form operation cost. The optimal control law can be derived from solving the Riccati Equation. We design an LQR controller by leveraging the continuous-time nominal model of the R2R system. The details of the LQR controller design is stated in Appendix B.

The purple line in Figure \ref{fig:Compare_PID} shows the performance of the LQR controller designed above. As observed with the PID controllers, the LQR controller similarly results in RE convergence to a non-zero steady-state value. Despite augmenting the state vector with integral terms in the LQR design, the controller remains insufficient to fully capture the complex RE dynamics and achieve complete elimination. This motivates us to add an additional STILC component to further improve the RE mitigation performance.

\subsection{Successful RE Elimination by Adding STILC Components}
\begin{figure} 
    \centering
    \includegraphics[width=0.48\textwidth]{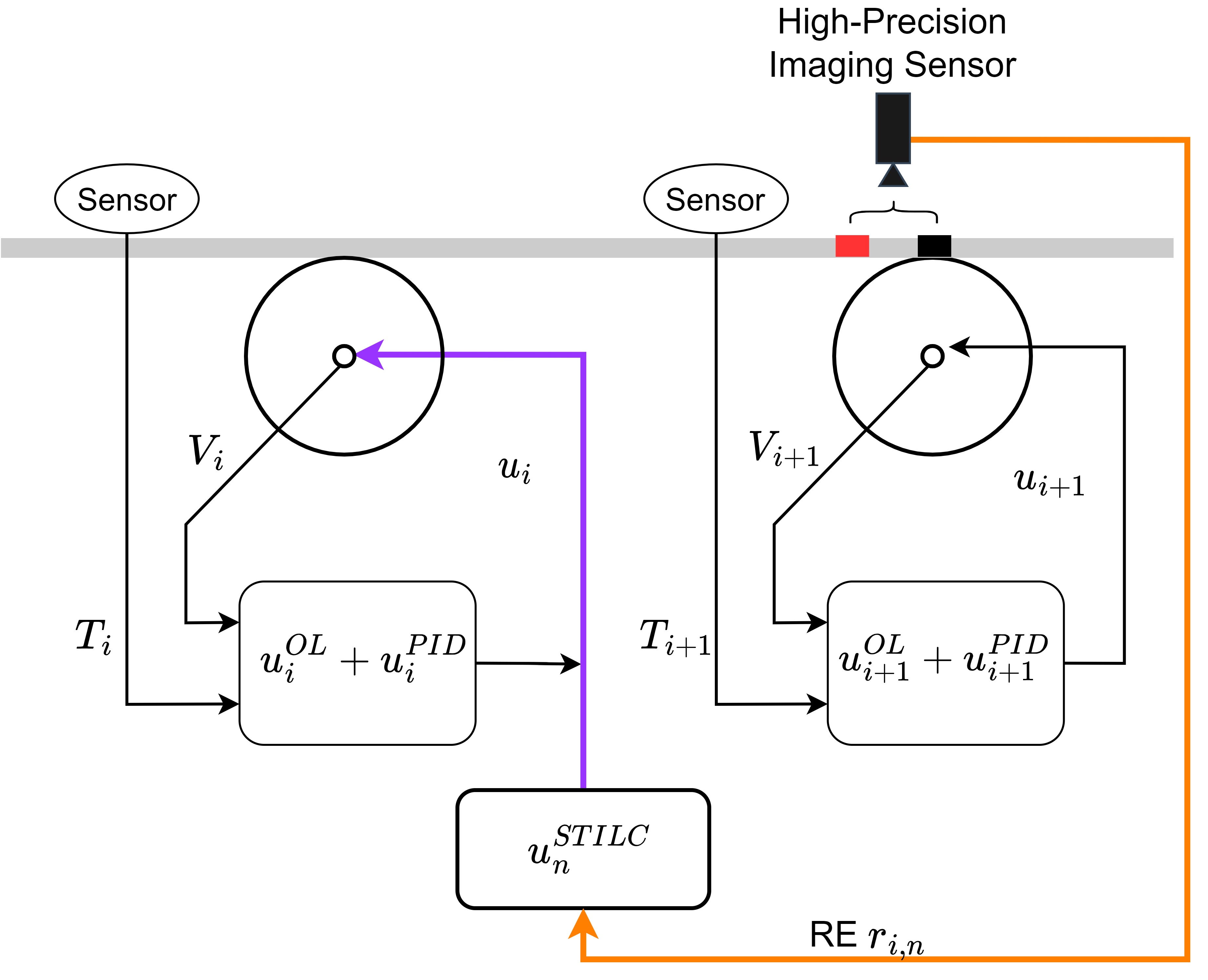}
    \caption{STILC-PID Hybrid Control for Registration Error}
    \label{fig:STILC}
\end{figure}

\begin{figure*}
\centering
\begin{subfigure}[b]{0.45\textwidth}
    \centering
    \includegraphics[width=\textwidth]{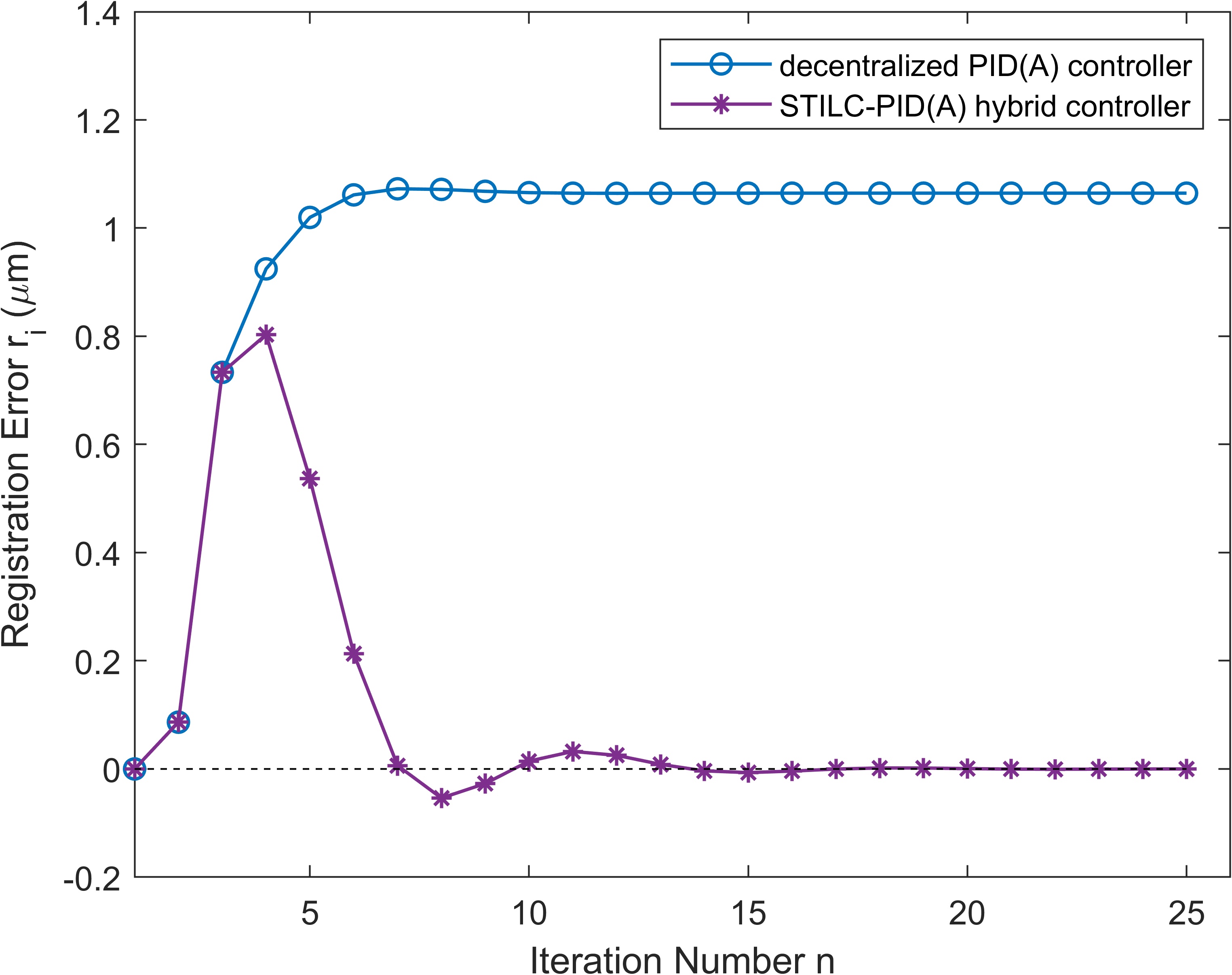}
    \caption{}
    \label{fig:Compare_controllers_A}
\end{subfigure}
\hfill
\begin{subfigure}[b]{0.45\textwidth}
    \centering
    \includegraphics[width=\textwidth]{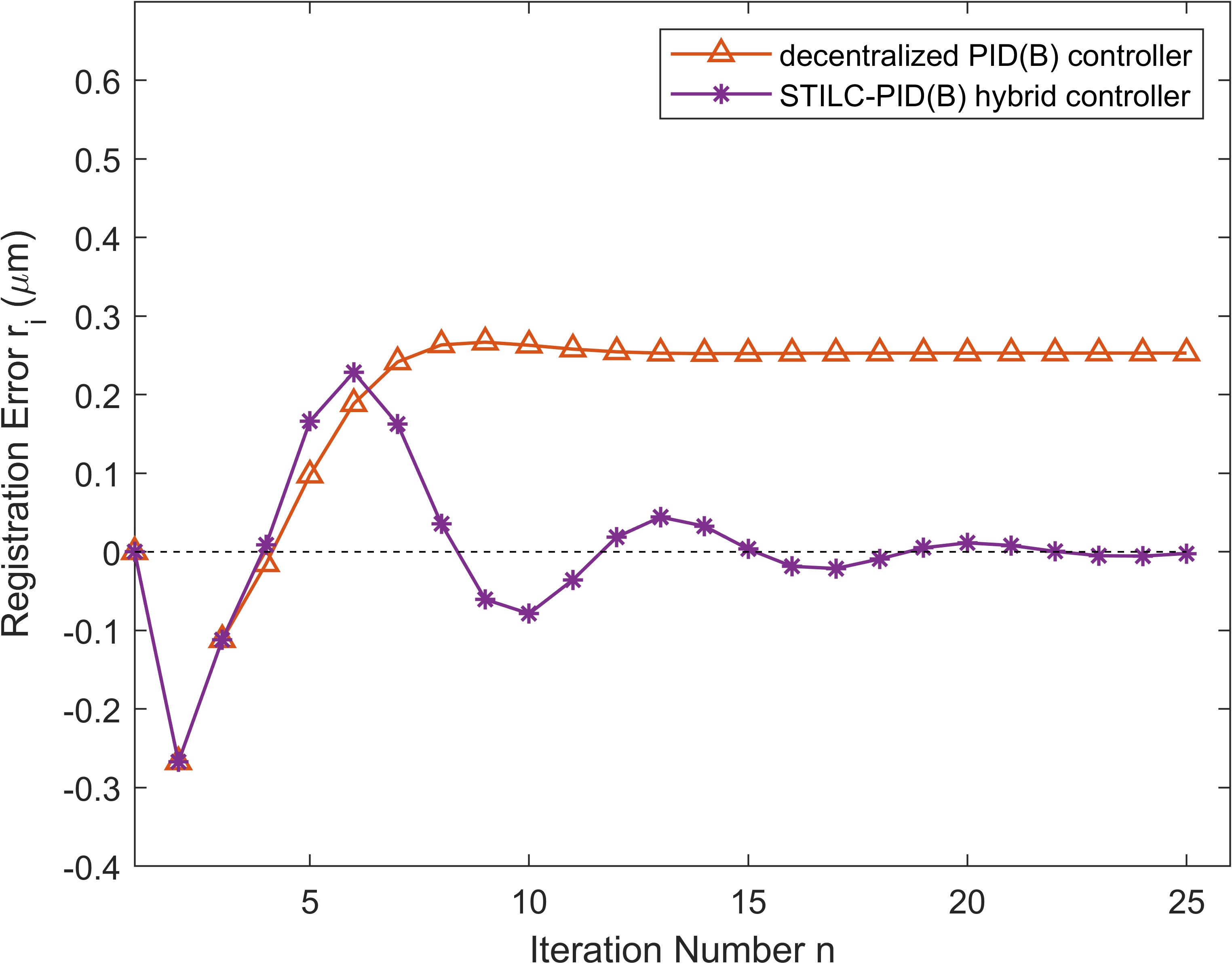}
    \caption{}
    \label{fig:Compare_controllers_B}
\end{subfigure}

\vspace{0.5cm}

\begin{subfigure}[b]{0.45\textwidth}
    \centering
    \includegraphics[width=\textwidth]{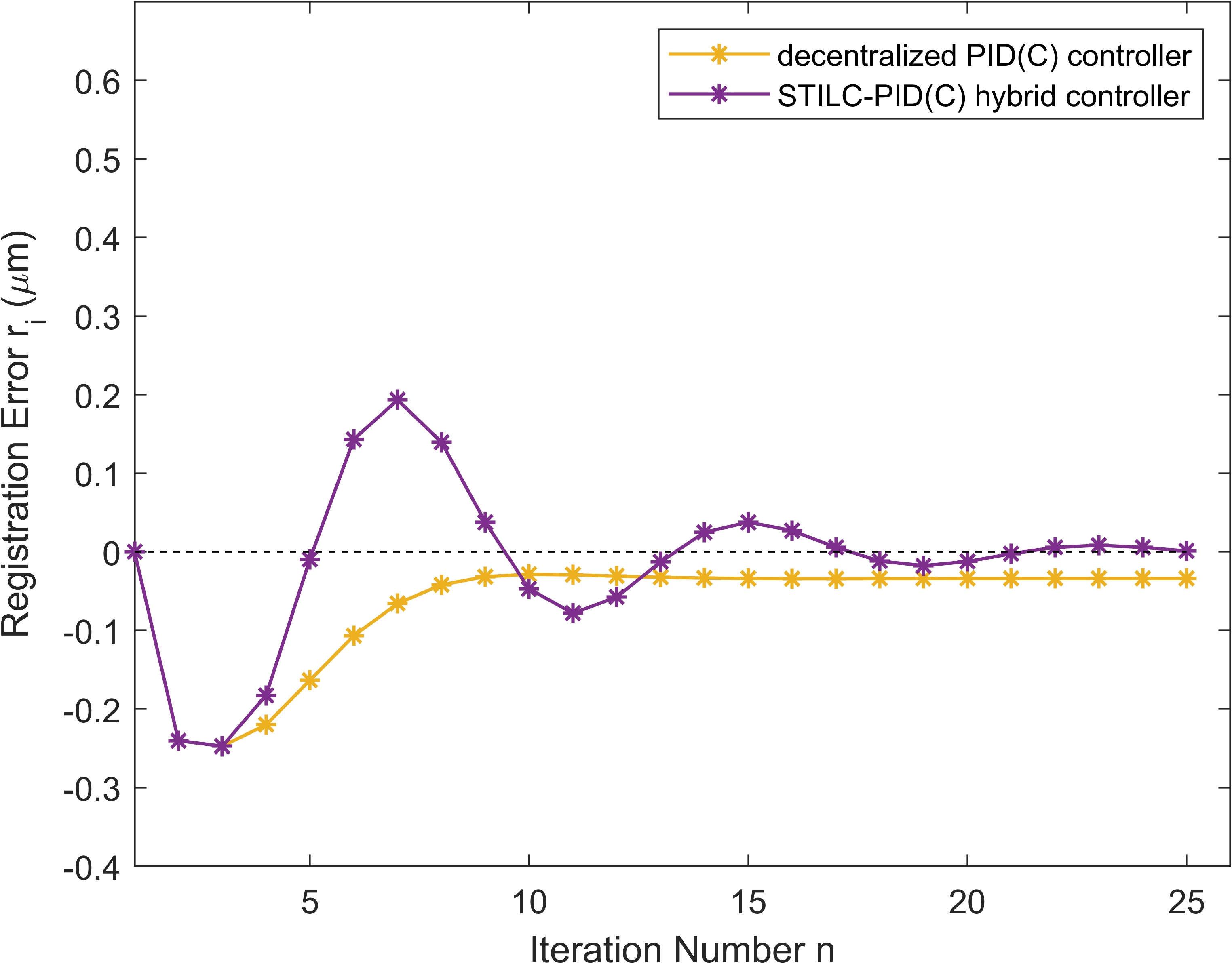}
    \caption{}
    \label{fig:Compare_controllers_C}
\end{subfigure}
\hfill
\begin{subfigure}[b]{0.45\textwidth}
    \centering
    \includegraphics[width=\textwidth]{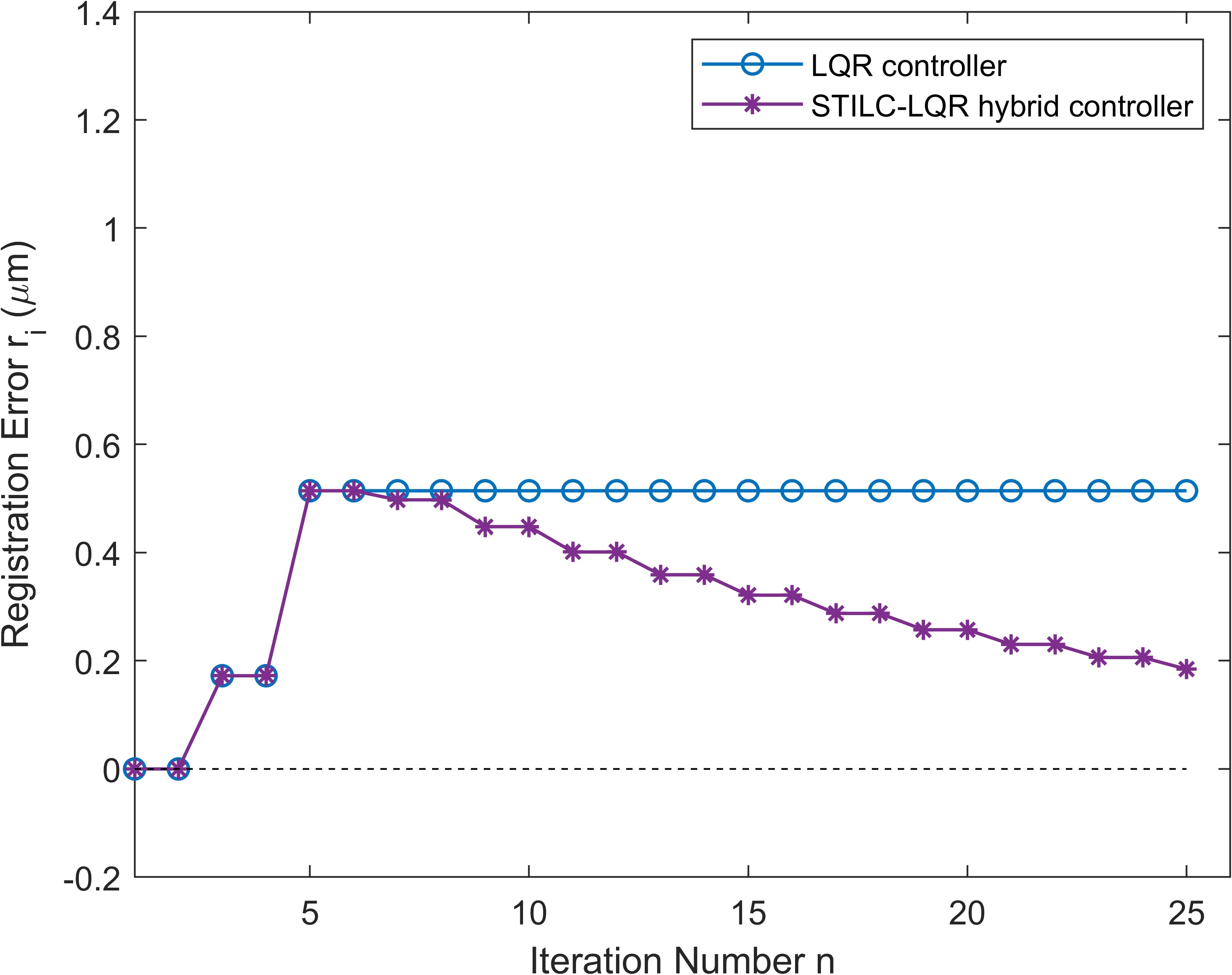}
    \caption{}
    \label{fig:Compare_controllers_LQR}
\end{subfigure}
\caption{Performance Comparison between Proposed STILC-based Hybrid Controller and the pure feedback controllers Controllers (a) PID A; (b) PID B; (c) PID C; (d) LQR}
\label{fig:Compare_controllers}
\end{figure*}

To guarantee that the registration converges to zero, we add a STILC component to the control input of the upstream roller, as shown in Figure \ref{fig:STILC}. As the target of this work is to propose a simplest pipeline of designing effective controllers to eliminate RE, we select PID as the baseline controller which requires much less expert knowledge than designing a LQR controller. Figure \ref{fig:Compare_controllers} shows the comparison between the registration control performances between the STILC-PID hybrid controller and the pure decentralized PID controllers with different PID parameter settings. The specific controller settings are given in Table \ref{tab:Controller_setting}. The simulation results show that the proposed hybrid controller with an additional STILC input component makes the registration error converge to zero after sufficient iterations (10-20 iterations in this simulation case). For PID A, the worst PID parameter setting in these three settings, an additional STILC component significantly reduces the registration error in 10 iterations and makes the error converge to zero, as shown in Figure \ref{fig:Compare_controllers_A}. Similar effects can be observed for PID B and PID C settings in Figure \ref{fig:Compare_controllers_B} and Figure \ref{fig:Compare_controllers_C}. For PID C, even though the purely decentralized PID achieves a relatively small registration error, the STILC-PID hybrid controller enhances the accuracy further. It completely eliminates the registration error after undergoing several oscillations in the iteration domain. It is worth noting that the added STILC is also effective for LQR controller, as observed in Figure \ref{fig:Compare_controllers_LQR}, which implies the generalizability of the STILC component. Therefore, the proposed hybrid controller shows a significant advantage in R2R printing registration control.

The control law of the hybrid controller is given as follows:

\begin{equation}\label{eq:8}
u_j(\tau)=u_j^{OL}+u_j^{PID}(\tau)+u_n^{STILC}(\theta_i(\tau))
\end{equation}

\subsection{Comparison of the Effects of Different Learning Gains}
\begin{figure} 
    \centering
    \includegraphics[width=0.45\textwidth]{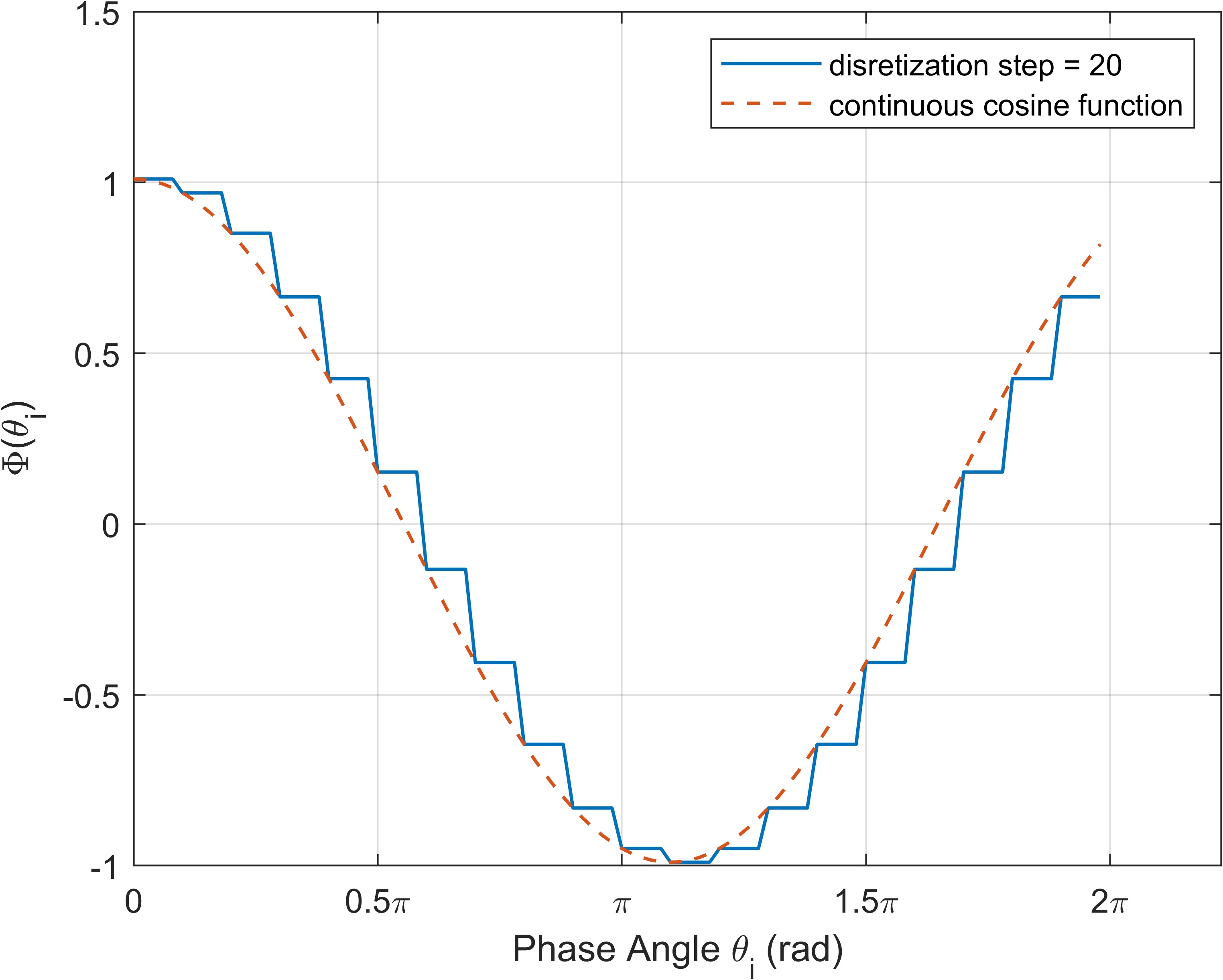}
    \caption{20-Step Discretized Cosine-Form Basis Function}
    \label{fig:discretized basis}
\end{figure}

\begin{figure} 
    \centering
    \includegraphics[width=0.45\textwidth]{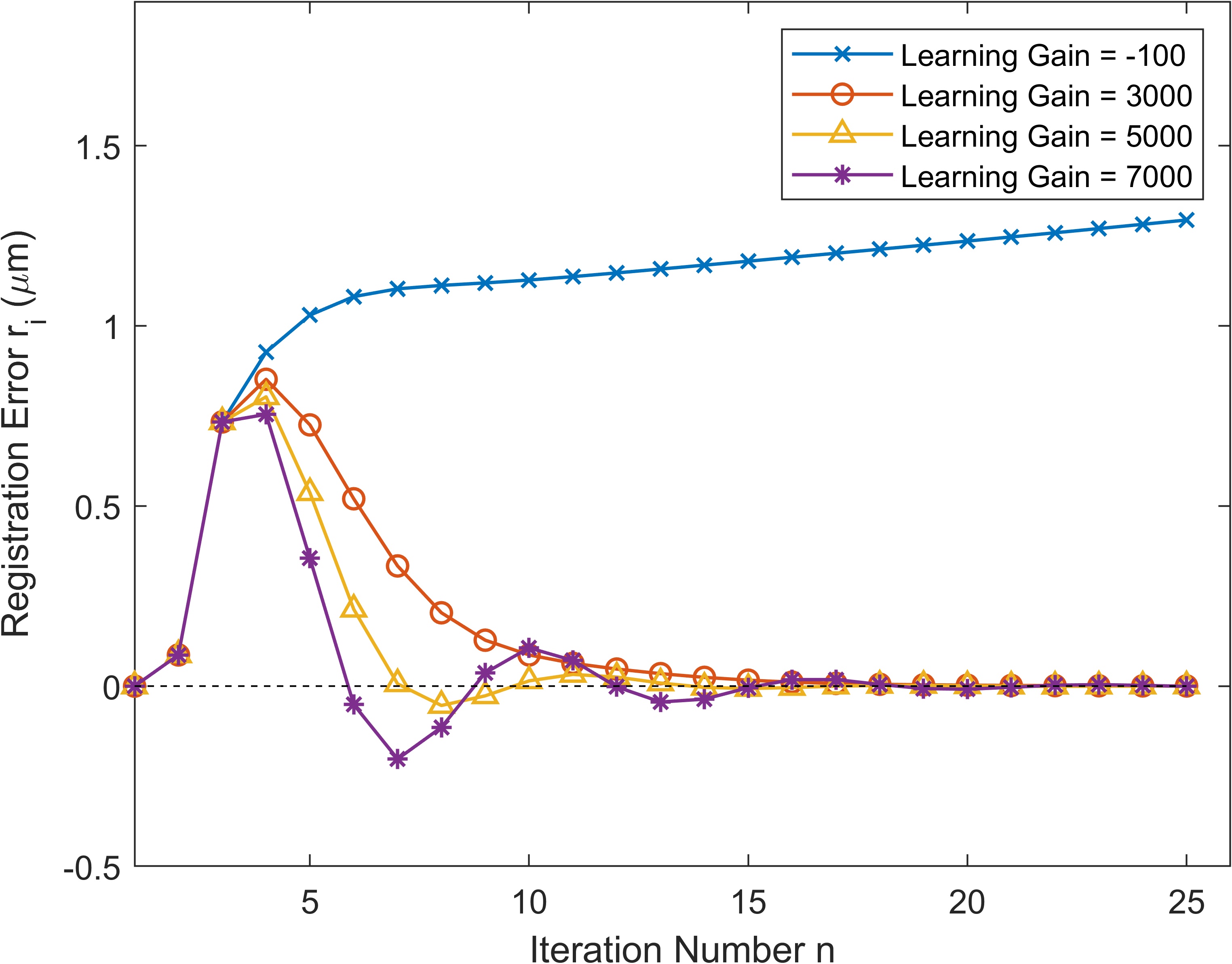}
    \caption{Learning Gain Effects Comparison}
    \label{fig:Learning Gain Compare}
\end{figure}

To make the STILC more feasible in real-world scenarios, we design a discretized basis function as shown in Fig.~\ref{fig:discretized basis}. The discretized basis function can be written in the memory of the controller hardware as a small lookup table. Thus, it can help avoid computing the cosine function in real-time. Figure \ref{fig:Compare_controllers} demonstrates that a 20-step discretized cosine-form basis function adequately approximates the disturbance profile $u_{dist}$, enabling the registration error to iteratively converge to zero, as we discuss in Eq.~\eqref{eq:ideal_Phi_L}. We also discuss the influence of learning gain selection in Section \ref{sec:3}. Figure \ref{fig:Learning Gain Compare} compares the effects of selecting different learning gains. It shows that a negative learning gain (-100) will make the registration error diverge, while positive learning gains (3000, 5000, 7000) can make the registration error converge to zero. Larger learning gains give us a faster response in the iteration domain and reduce the registration error more aggressively, but also cause larger overshoot and fluctuations. By numerical computation, we know selecting $\mathcal{L}=5000$ approximately satisfies $(\Omega_1-1)^2+4\Omega_2 = 0$. And the simulation result shows $\mathcal{L}=5000$ can provide balanced performance for convergence speed and fluctuation. A smaller learning gain can be selected if a monotonic and asymptotic convergence is required, such as $\mathcal{L}=3000$. These results have verified our analysis of the learning gain selection principle at the end of Section \ref{sec:3}.

\begin{table}[!t]
\caption[Table]{PID and STILC Settings\label{tab:Controller_setting}}
\scalebox{0.85}{
\begin{tabular}{lcc}
\hline
Setting & Notation & Value \\
\hline
PID Gain (P) & $K^P_i$,$K^P_{i+1}$ & $[-0.1916 ~~~~~ 0]$ \\
PID Gain (I) & $K^I_i$,$K^I_{i+1}$ & $[-0.0767 ~~~~~ 0]$ \\
PID Gain (D) & $K^D_i$,$K^D_{i+1}$ & $[-0.0038 ~ -0.1916]$ \\
ILC Learning Gain & $P$ & 5000 \\
Basis Function & $G$ & Shown in Fig.~\ref{fig:discretized basis} (20-step)\\
\hline
\end{tabular}
}
\end{table}

\section{Conclusion}
\label{sec:5}
This paper presents a novel Spatial-Terminal Iterative Learning Control (STILC)-PID hybrid control method to address a fundamental challenge in registration control within Roll-to-Roll (R2R) printing systems: the inability to monitor real-time registration error (RE). The proposed method overcomes this limitation by incorporating a Terminal Iterative Learning Control (TILC) updating law with a spatially dependent basis function, enabling iterative convergence of the RE to zero. This approach contrasts with PID type of feedback control methods, which can only achieve convergence to a non-zero level.

Our work builds upon the concept of Spatial Iterative Learning Control (SILC), adapting it specifically for R2R printing processes. We leverage the insight that rollers in R2R systems exhibit spatially (angularly) periodic behavior, a characteristic common to rotary machinery. This allows us to design a cosine-form basis function in the spatial domain, forming the foundation of our STILC method.
The paper derives controller design criteria to ensure the convergence of the proposed STILC-PID hybrid control approach. These criteria provide clear guidance for practitioners to design the basis function and select the proper learning gain, significantly simplifying the controller design process. This feature is particularly beneficial for broadening the applicability of STILC to various repetitive manufacturing processes governed by linear time-varying (LTV) dynamics.

To evaluate the effectiveness of our method, we apply it to a registration control problem in R2R printing systems. We employ the perturbation method to model the dynamics of the RE and the R2R system. Additionally, we introduce a common source of disturbance—a spatially-dependent axis mismatch between the motor shaft and the roller center. Through simulation experiments conducted in Simulink, we demonstrate the effectiveness of the proposed STILC-PID hybrid control method. The simulation results clearly show that our method achieves convergence of the RE to zero in the iteration domain, while the traditional decentralized PID method only achieves convergence to a non-zero value.
Our STILC method offers several key advantages over existing approaches:

(1) It is a learning-based data-driven control method, minimizing requirements on designers' expertise and hardware computation capabilities.

(2) It updates the control input profile by adjusting only one parameter based on the terminal RE measurement, significantly reducing the computational burden for real-time control.

(3) It provides a highly simplified approach to controller design, enhancing its accessibility to practitioners across various applications while maintaining its effectiveness.

As part of future work, we plan to enhance our method to handle practical disturbances encountered in R2R printing systems, such as roller roundness errors and axis mismatches with varying initial phase angles. Moreover, we intend to validate the proposed method in industrial application scenarios involving general rotary machine systems, which inherently exhibit angle-periodic behaviors.

Our proposed STILC method holds significant promise as a solution to widespread industrial problems since angle-periodic behaviors broadly exist in various machines driven by motors. For example, in multi-axis motion systems, such as those used in Computer Numerical Control (CNC) machining, advanced robotics, and 3D printing, the tool paths or actuator trajectories are essentially determined by the angle-periodic behaviors of drive motors. Our method presents a suitable solution to compensate for angle-dependent repetitive disturbances and help achieve better precision across these diverse applications.


\section*{Acknowledgments}
This material is based upon work supported by the National Science Foundation under grant CMMI-1943801.


\section*{Appendix A: Proof of Theorem 1}
By \eqref{eq:E_eq_1} we can derive

\begin{equation}\label{eq:E_eq_2}
E_{j+1} = \Omega_1 E_{j} + \Omega_2 \sum_{s=1}^{j-1} E_s +\Omega_3.
\end{equation}

Since the system \eqref{eq:dt_stable_sys} is stabilized, we can assume the initial state $x_j(0)$ for any iteration $j$ is identical. Subtracting \eqref{eq:E_eq_1} from \eqref{eq:E_eq_2} yields

\begin{equation}\label{eq:E_eq_d}
E_{j+1}-E_{j} = \Omega_1 E_{j} - \Omega_1 E_{j-1} +\Omega_2 E_{j-1}.
\end{equation}

Rearranging (32), we obtain

\begin{equation}\label{eq:E_eq_recurrence}
E_{j+1}-(\Omega_1+1)E_{j}+(\Omega_1-\Omega_2)E_{j-1} = 0.
\end{equation}

Thus, the terminal error satisfies a second-order recurrence relation. We can obtain the general solution\cite{papanicolaou1996asymptotic} for the second-order recurrence relation  \eqref{eq:E_eq_recurrence}.

If $\lambda_1 \neq \lambda_2$, then the general solution is

\begin{equation}\label{eq:general_sol_unequal}
E_j=\xi_1 \lambda_1^j+\xi_2 \lambda_2^j.
\end{equation}

If $\lambda_1 = \lambda_2 = \lambda_0$, then the general solution is

\begin{equation}\label{eq:general_sol_equal}
E_j=(\xi_1+\xi_2 j)\lambda_0^j.
\end{equation}

$\xi_1$ and $\xi_2$ are two bounded complex numbers and can be solved when the first two errors $E_0$, $E_1$ are known.

From the general solutions \eqref{eq:general_sol_unequal} and \eqref{eq:general_sol_equal}, it is obvious that $\lim_{j\to\infty}E_j=0$ if $\lvert \lambda_1\rvert <1$ and $\lvert \lambda_2\rvert <1$.

$\blacksquare$

\section*{Appendix B: LQR Controller Design}
Because the RE is related to the integral of tensions (see \eqref{eq:6}), we augment the system with integral states:

\begin{equation}
\begin{bmatrix} \dot{x}(t) \\ \dot{z}(t) \end{bmatrix} = 
\begin{bmatrix} A_{nominal} & 0 \\ -C_{nominal} & 0 \end{bmatrix}
\begin{bmatrix} x(t) \\ z(t) \end{bmatrix} +
\begin{bmatrix} B_{nominal} \\ 0 \end{bmatrix} u(t)
\end{equation}

where $z(t)$ represents the integral of the states. The matrices of the nominal system are:

$$A_{nominal} = \begin{bmatrix}
-\frac{vr_1}{L_1} & 0 & 0 & \frac{AE-tr_1}{L_1} & 0 \\
\frac{vr_1}{L_2} & -\frac{vr_2}{L_2} & 0 & \frac{tr_1-AE}{L_2} & \frac{AE-tr_2}{L_2} \\
0 & \frac{vr_2}{L_3} & -\frac{vr3}{L_3} & 0 & \frac{tr_2-AE}{L_3} \\
-\frac{R_1^2}{J_1} & \frac{R_1^2}{J_1} & 0 & -\frac{bf_1}{J_1} & 0 \\
0 & -\frac{R_2^2}{J_2} & \frac{R_2^2}{J_2} & 0 & -\frac{bf_2}{J_2}
\end{bmatrix}$$

$$B_{nominal} = \begin{bmatrix}
0 & 0 \\
0 & 0 \\
0 & 0 \\
\frac{n_1R_1}{J_1} & 0 \\
0 & \frac{n_2R_2}{J_2}
\end{bmatrix}$$

$$C_{nominal} = \begin{bmatrix}
1 & 0 & 0 & 0 & 0 \\
0 & 1 & 0 & 0 & 0
\end{bmatrix}$$.

For the LQR design, we need the augmented system matrices

$$A_{aug} = \begin{bmatrix} A_{nominal} & 0 \\ -C & 0 \end{bmatrix}, \quad
B_{aug} = \begin{bmatrix} B_{nominal} \\ 0 \end{bmatrix}$$

and we design the cost function:

\begin{equation}
J = \int_0^\infty (x_{aug}^T Q x_{aug} + u^T R u) dt
\end{equation}

where $x_{aug} = [x \; z]^T$, and $Q$ and $R$ are positive definite weighting matrices. In the following simulation, we define $Q$ and $R$ as

$$Q = \begin{bmatrix}
10 & 0 & 0 & 0 & 0 & 0 & 0 \\
0 & 10 & 0 & 0 & 0 & 0 & 0 \\
0 & 0 & 1 & 0 & 0 & 0 & 0 \\
0 & 0 & 0 & 1 & 0 & 0 & 0 \\
0 & 0 & 0 & 0 & 1 & 0 & 0 \\
0 & 0 & 0 & 0 & 0 & 100 & 0 \\
0 & 0 & 0 & 0 & 0 & 0 & 100
\end{bmatrix}$$

$$R = \begin{bmatrix}
5 & 0 \\
0 & 5
\end{bmatrix}$$

Then we solve the algebraic Riccati equation:

\begin{equation}
\label{eq:Riccati}
A_{aug}^T P + PA_{aug} - PB_{aug}R^{-1}B_{aug}^T P + Q = 0    
\end{equation}

The positive definite matrix $P$ solved from \eqref{eq:Riccati} can be used to calculate the optimal feedback gain $K_{LQR} = R^{-1}B_{aug}^T P$. The final control law is given by:

\begin{equation}
u(t) = -K_{LQR} \begin{bmatrix} x(t) \\ z(t) \end{bmatrix}
\end{equation}

\bibliographystyle{IEEEtran}
\bibliography{STILC_R2R}

\begin{thebibliography}{10}
\providecommand{\url}[1]{#1}
\csname url@samestyle\endcsname
\providecommand{\newblock}{\relax}
\providecommand{\bibinfo}[2]{#2}
\providecommand{\BIBentrySTDinterwordspacing}{\spaceskip=0pt\relax}
\providecommand{\BIBentryALTinterwordstretchfactor}{4}
\providecommand{\BIBentryALTinterwordspacing}{\spaceskip=\fontdimen2\font plus
\BIBentryALTinterwordstretchfactor\fontdimen3\font minus \fontdimen4\font\relax}
\providecommand{\BIBforeignlanguage}[2]{{%
\expandafter\ifx\csname l@#1\endcsname\relax
\typeout{** WARNING: IEEEtran.bst: No hyphenation pattern has been}%
\typeout{** loaded for the language `#1'. Using the pattern for}%
\typeout{** the default language instead.}%
\else
\language=\csname l@#1\endcsname
\fi
#2}}
\providecommand{\BIBdecl}{\relax}
\BIBdecl

\bibitem{chen2019towards}
C.-M. Chen, S.~Anastasova, K.~Zhang, B.~G. Rosa, B.~P. Lo, H.~E. Assender, and G.-Z. Yang, ``Towards wearable and flexible sensors and circuits integration for stress monitoring,'' \emph{IEEE Journal of Biomedical and Health Informatics}, vol.~24, no.~8, pp. 2208--2215, 2019.

\bibitem{kwon2018development}
S.~Kwon, D.~Song, H.~Kim, M.~Lee, and K.~Woo, ``Development of roll-to-roll multi-layer thermal evaporation system for flexible oled devices,'' in \emph{2018 25th International Workshop on Active-Matrix Flatpanel Displays and Devices (AM-FPD)}.\hskip 1em plus 0.5em minus 0.4em\relax IEEE, 2018, pp. 1--2.

\bibitem{wood2020perspectives}
D.~L. Wood~III, M.~Wood, J.~Li, Z.~Du, R.~E. Ruther, K.~A. Hays, N.~Muralidharan, L.~Geng, C.~Mao, and I.~Belharouak, ``Perspectives on the relationship between materials chemistry and roll-to-roll electrode manufacturing for high-energy lithium-ion batteries,'' \emph{Energy Storage Materials}, vol.~29, pp. 254--265, 2020.

\bibitem{di2018large}
F.~Di~Giacomo, H.~Fledderus, H.~Gorter, G.~Kirchner, I.~de~Vries, I.~Dogan, W.~Verhees, V.~Zardetto, M.~Najafi, D.~Zhang \emph{et~al.}, ``Large area> 140 cm 2 perovskite solar modules made by sheet to sheet and roll to roll fabrication with 14.5\% efficiency,'' in \emph{2018 IEEE 7th world conference on photovoltaic energy conversion (WCPEC)(a joint conference of 45th IEEE PVSC, 28th PVSEC \& 34th EU PVSEC)}.\hskip 1em plus 0.5em minus 0.4em\relax IEEE, 2018, pp. 2795--2798.

\bibitem{salari2019investigation}
M.~Salari and M.~Joodaki, ``Investigation of electrical characteristics dependency of roll-to-roll printed solar cells with silver electrodes on mechanical tensile strain,'' \emph{IEEE Transactions on Device and Materials Reliability}, vol.~19, no.~4, pp. 718--722, 2019.

\bibitem{yang2024roll}
M.~Yang, H.~Li, J.~Wang, W.~Shi, Q.~Zhang, H.~Xing, W.~Ren, B.~Sun, M.~Guo, E.~Xu \emph{et~al.}, ``Roll-to-roll fabricated polymer composites filled with subnanosheets exhibiting high energy density and cyclic stability at 200Â° c,'' \emph{Nature Energy}, vol.~9, no.~2, pp. 143--153, 2024.

\bibitem{parvazian2024roll}
E.~Parvazian and T.~Watson, ``The roll-to-roll revolution to tackle the industrial leap for perovskite solar cells,'' \emph{nature communications}, vol.~15, no.~1, p. 3983, 2024.

\bibitem{valimaki2022accuracy}
M.~K. V{\"a}lim{\"a}ki, E.~Jansson, V.~J. Von~Morgen, M.~Ylikunnari, K.-L. V{\"a}is{\"a}nen, P.~Ontero, M.~Kehusmaa, P.~Korhonen, and T.~M. Kraft, ``Accuracy control for roll and sheet processed printed electronics on flexible plastic substrates,'' \emph{The International Journal of Advanced Manufacturing Technology}, vol. 119, no.~9, pp. 6255--6273, 2022.

\bibitem{chen2020fully}
Z.~Chen, T.~Zhang, Y.~Zheng, D.~S.-H. Wong, and Z.~Deng, ``Fully decoupled control of the machine directional register in roll-to-roll printing system,'' \emph{IEEE Transactions on Industrial Electronics}, vol.~68, no.~10, pp. 10\,007--10\,018, 2020.

\bibitem{chen2018modeling}
Z.~Chen, Y.~Zheng, T.~Zhang, D.~S.-H. Wong, and Z.~Deng, ``Modeling and register control of the speed-up phase in roll-to-roll printing systems,'' \emph{IEEE Transactions on Automation Science and Engineering}, vol.~16, no.~3, pp. 1438--1449, 2018.

\bibitem{kang2013modeling}
H.~Kang, C.~Lee, and K.~Shin, ``Modeling and compensation of the machine directional register in roll-to-roll printing,'' \emph{Control Engineering Practice}, vol.~21, no.~5, pp. 645--654, 2013.

\bibitem{CHEN2023651}
Z.~Chen, B.~Qu, B.~Jiang, S.~R. Forrest, and J.~Ni, ``Robust constrained tension control for high-precision roll-to-roll processes,'' \emph{ISA Transactions}, vol. 136, pp. 651--662, 2023.

\bibitem{shah2022data}
K.~Shah, A.~He, Z.~Wang, X.~Du, and X.~Jin, ``Data-driven model predictive control for roll-to-roll process register error,'' in \emph{International Manufacturing Science and Engineering Conference}, vol. 86601.\hskip 1em plus 0.5em minus 0.4em\relax American Society of Mechanical Engineers, 2022, p. V001T03A006.

\bibitem{pagilla2006decentralized}
P.~R. Pagilla, N.~B. Siraskar, and R.~V. Dwivedula, ``Decentralized control of web processing lines,'' \emph{IEEE Transactions on control systems technology}, vol.~15, no.~1, pp. 106--117, 2006.

\bibitem{zhou2014model}
M.~Zhou, Z.~Chen, Y.~Zheng, and L.~Zou, ``Model based pd control during the speed-up phase in roll-to-roll web register systems,'' in \emph{Proceedings of the 33rd Chinese Control Conference}.\hskip 1em plus 0.5em minus 0.4em\relax IEEE, 2014, pp. 5139--5143.

\bibitem{chen2016model}
Z.~Chen, Y.~Zheng, M.~Zhou, D.~S.-H. Wong, L.~Chen, and Z.~Deng, ``Model-based feedforward register control of roll-to-roll web printing systems,'' \emph{Control Engineering Practice}, vol.~51, pp. 58--68, 2016.

\bibitem{bristow2006survey}
D.~A. Bristow, M.~Tharayil, and A.~G. Alleyne, ``A survey of iterative learning control,'' \emph{IEEE control systems magazine}, vol.~26, no.~3, pp. 96--114, 2006.

\bibitem{lee2007iterative}
J.~H. Lee and K.~S. Lee, ``Iterative learning control applied to batch processes: An overview,'' \emph{Control Engineering Practice}, vol.~15, no.~10, pp. 1306--1318, 2007.

\bibitem{sutanto2013norm}
E.~Sutanto and A.~G. Alleyne, ``Norm optimal iterative learning control for a roll to roll nano/micro-manufacturing system,'' in \emph{2013 American Control Conference}.\hskip 1em plus 0.5em minus 0.4em\relax IEEE, 2013, pp. 5935--5941.

\bibitem{sutanto2014vision}
------, ``Vision based iterative learning control for a roll to roll micro/nano-manufacturing system,'' \emph{IFAC Proceedings Volumes}, vol.~47, no.~3, pp. 7202--7207, 2014.

\bibitem{kim2017measurement}
C.~Kim, S.~W. Jeon, and C.~H. Kim, ``Measurement of position accuracy of engraving in plate roller and its effect on register accuracy in roll-to-roll multi-layer printing,'' \emph{Measurement Science and Technology}, vol.~28, no.~12, p. 125002, 2017.

\bibitem{lee2015register}
J.~Lee, J.~Seong, J.~Park, S.~Park, D.~Lee, and K.-H. Shin, ``Register control algorithm for high resolution multilayer printing in the roll-to-roll process,'' \emph{Mechanical Systems and Signal Processing}, vol.~60, pp. 706--714, 2015.

\bibitem{lee2018smearing}
J.~Lee, S.~Park, K.-H. Shin, and H.~Jung, ``Smearing defects: a root cause of register measurement error in roll-to-roll additive manufacturing system,'' \emph{The International Journal of Advanced Manufacturing Technology}, vol.~98, no.~9, pp. 3155--3165, 2018.

\bibitem{arimoto1984bettering}
S.~Arimoto, S.~Kawamura, and F.~Miyazaki, ``Bettering operation of robots by learning,'' \emph{Journal of Robotic systems}, vol.~1, no.~2, pp. 123--140, 1984.

\bibitem{4048052}
------, ``Bettering operation of dynamic systems by learning: A new control theory for servomechanism or mechatronics systems,'' in \emph{The 23rd IEEE Conference on Decision and Control}, 1984, pp. 1064--1069.

\bibitem{han2018terminal}
J.~Han, D.~Shen, and C.-J. Chien, ``Terminal iterative learning control for discrete-time nonlinear systems based on neural networks,'' \emph{Journal of the Franklin Institute}, vol. 355, no.~8, pp. 3641--3658, 2018.

\bibitem{wang2024adaptive}
Z.~Wang and X.~Jin, ``An adaptive spatial-terminal iterative learning strategies in roll-to-roll control problems,'' in \emph{ASME International Mechanical Engineering Congress and Exposition}, vol. 88605.\hskip 1em plus 0.5em minus 0.4em\relax American Society of Mechanical Engineers, 2024, p. V002T03A104.

\bibitem{wang2025adaptive}
------, ``An adaptive spatial-terminal iterative learning strategy in roll-to-roll control problems,'' \emph{Journal of Micro and Nano Science and Engineering}, vol.~13, no.~3, p. 034502, 2025.

\bibitem{xu1999terminal}
J.-X. Xu, Y.~Chen, T.~H. Lee, and S.~Yamamoto, ``Terminal iterative learning control with an application to rtpcvd thickness control,'' \emph{Automatica}, vol.~35, no.~9, pp. 1535--1542, 1999.

\bibitem{bu2020data}
X.~Bu, J.~Liang, Z.~Hou, and R.~Chi, ``Data-driven terminal iterative learning consensus for nonlinear multiagent systems with output saturation,'' \emph{IEEE transactions on neural networks and learning systems}, vol.~32, no.~5, pp. 1963--1973, 2020.

\bibitem{chi2012data}
R.~Chi, D.~Wang, Z.~Hou, and S.~Jin, ``Data-driven optimal terminal iterative learning control,'' \emph{Journal of Process Control}, vol.~22, no.~10, pp. 2026--2037, 2012.

\bibitem{xu2008spatial}
J.-X. Xu and D.~Huang, ``Spatial periodic adaptive control for rotary machine systems,'' \emph{IEEE Transactions on Automatic Control}, vol.~53, no.~10, pp. 2402--2408, 2008.

\bibitem{AfkhamiZahra2021SILC}
Z.~Afkhami, C.~Pannier, L.~Aarnoudse, D.~Hoelzle, and K.~Barton, ``\BIBforeignlanguage{eng}{Spatial iterative learning control for multi-material three-dimensional structures},'' \emph{\BIBforeignlanguage{eng}{ASME letters in dynamic systems and control}}, vol.~1, no.~1, 2021.

\bibitem{afkhami2023robust}
Z.~Afkhami, D.~J. Hoelzle, and K.~Barton, ``Robust higher-order spatial iterative learning control for additive manufacturing systems,'' \emph{IEEE Transactions on Control Systems Technology}, 2023.

\bibitem{afkhami2021higher}
Z.~Afkhami, D.~Hoelzle, and K.~Barton, ``Higher-order spatial iterative learning control for additive manufacturing,'' in \emph{2021 60th IEEE Conference on Decision and Control (CDC)}.\hskip 1em plus 0.5em minus 0.4em\relax IEEE, 2021, pp. 6547--6553.

\bibitem{LiuYan2018SILC}
Y.~Liu and X.~Ruan, ``\BIBforeignlanguage{eng}{Spatial iterative learning control for pitch of wind turbine},'' in \emph{\BIBforeignlanguage{eng}{2018 IEEE 7th Data Driven Control and Learning Systems Conference (DDCLS)}}.\hskip 1em plus 0.5em minus 0.4em\relax IEEE, 2018, pp. 841--846.

\bibitem{yang2022spatial}
L.~Yang, Y.~Li, D.~Huang, J.~Xia, and X.~Zhou, ``Spatial iterative learning control for robotic path learning,'' \emph{IEEE Transactions on Cybernetics}, 2022.

\bibitem{li2022five}
J.~Li, Z.~You, Y.~Li, E.~Miao, and R.~Yue, ``Five-axis contour error control based on spatial iterative learning,'' \emph{IEEE Transactions on Automation Science and Engineering}, vol.~20, no.~1, pp. 112--123, 2022.

\bibitem{li2021constrained}
Z.~Li, C.~Yin, H.~Ji, and Z.~Hou, ``Constrained spatial adaptive iterative learning control for trajectory tracking of high speed train,'' \emph{IEEE Transactions on Intelligent Transportation Systems}, vol.~23, no.~8, pp. 11\,720--11\,728, 2021.

\bibitem{li2020spatial}
Z.~Li, Z.~Hou, and C.~Yin, ``Spatial adaptive iterative learning control for automatic driving of high speed train,'' in \emph{2020 IEEE 9th Data Driven Control and Learning Systems Conference (DDCLS)}.\hskip 1em plus 0.5em minus 0.4em\relax IEEE, 2020, pp. 1440--1445.

\bibitem{zhu2023spatial}
Z.~Zhu and X.~Zhang, ``Spatial adaptive iterative learning control for high-speed train with unknown speed delays,'' \emph{Proceedings of the Institution of Mechanical Engineers, Part I: Journal of Systems and Control Engineering}, p. 09596518231155960, 2023.

\bibitem{huang2023spatial}
D.~Huang, Y.~He, W.~Yu, N.~Qin, Q.~Wang, and P.~Sun, ``Spatial adaptive iterative learning tracking control for high-speed trains considering passing through neutral sections,'' \emph{IEEE Transactions on Systems, Man, and Cybernetics: Systems}, 2023.

\bibitem{xin2023spatial}
Z.~Xin and Z.~Zijun, ``Spatial adaptive iterative learning control based high-speed train operation tracking under external disturbance,'' \emph{Automatic Control and Computer Sciences}, vol.~57, no.~3, pp. 276--286, 2023.

\bibitem{kim2023terminal}
B.~Kim and S.~B. Choi, ``Terminal iterative learning control for an electrical powertrain system with backlash,'' in \emph{2023 IEEE International Conference on Systems, Man, and Cybernetics (SMC)}.\hskip 1em plus 0.5em minus 0.4em\relax IEEE, 2023, pp. 2991--2996.

\bibitem{wang2023spatial}
Z.~Wang and X.~Jin, ``Spatial-terminal iterative learning control for registration error elimination in high-precision roll-to-roll printing systems,'' in \emph{International Manufacturing Science and Engineering Conference}, vol. 87240.\hskip 1em plus 0.5em minus 0.4em\relax American Society of Mechanical Engineers, 2023, p. V002T05A009.

\bibitem{cobos2018spatially}
E.~O. Cobos~Torres and P.~R. Pagilla, ``Spatially dependent transfer functions for web lateral dynamics in roll-to-roll manufacturing,'' \emph{Journal of Dynamic Systems, Measurement, and Control}, vol. 140, no.~11, p. 111011, 2018.

\bibitem{hoelzle2015spatial}
D.~J. Hoelzle and K.~L. Barton, ``On spatial iterative learning control via 2-d convolution: Stability analysis and computational efficiency,'' \emph{IEEE Transactions on Control Systems Technology}, vol.~24, no.~4, pp. 1504--1512, 2015.

\bibitem{papanicolaou1996asymptotic}
V.~G. Papanicolaou, ``On the asymptotic stability of a class of linear difference equations,'' \emph{Mathematics Magazine}, vol.~69, no.~1, pp. 34--43, 1996.

\end{thebibliography}

\begin{IEEEbiography}[{\includegraphics[width=1in,height=1.25in,clip,keepaspectratio]{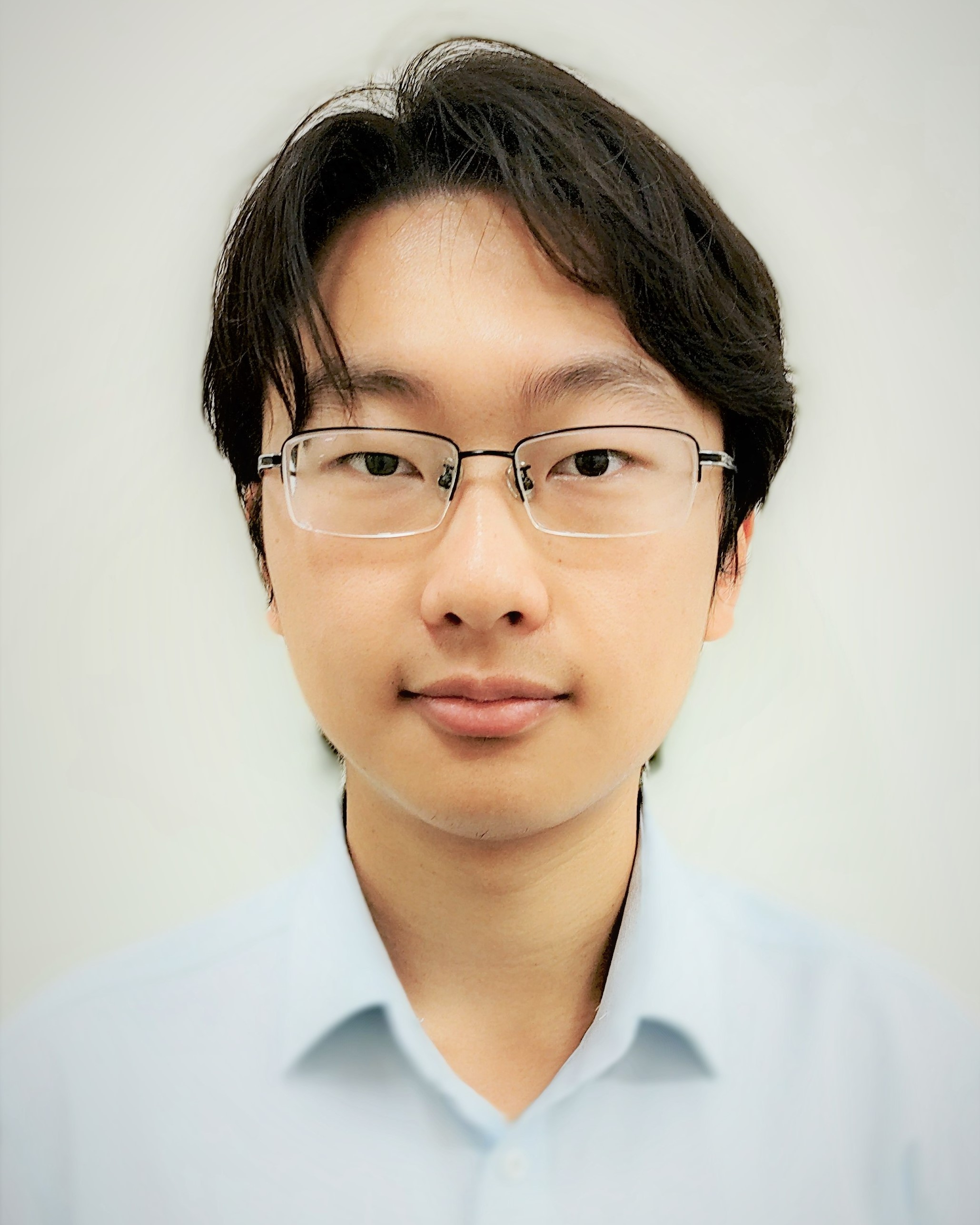}}]{Zifeng Wang} received his B.S degree in mechanical engineering from Shanghai Jiao Tong University in China in 2018. He was a mechanical engineer in Beko China R\&D Center from 2018 to 2020. He is currently a Ph.D candidate in Industrial Engineering at Northeastern University, Boston, MA, USA. His research interests include online learning control and data-driven methods in advanced manufacturing processes, aimed at leveraging the repetitiveness and abundant data in manufacturing scenarios to improve the automation and intelligence of industrial production systems.
 
\end{IEEEbiography}

\begin{IEEEbiography}[{\includegraphics[width=1in,height=1.25in,clip,keepaspectratio]{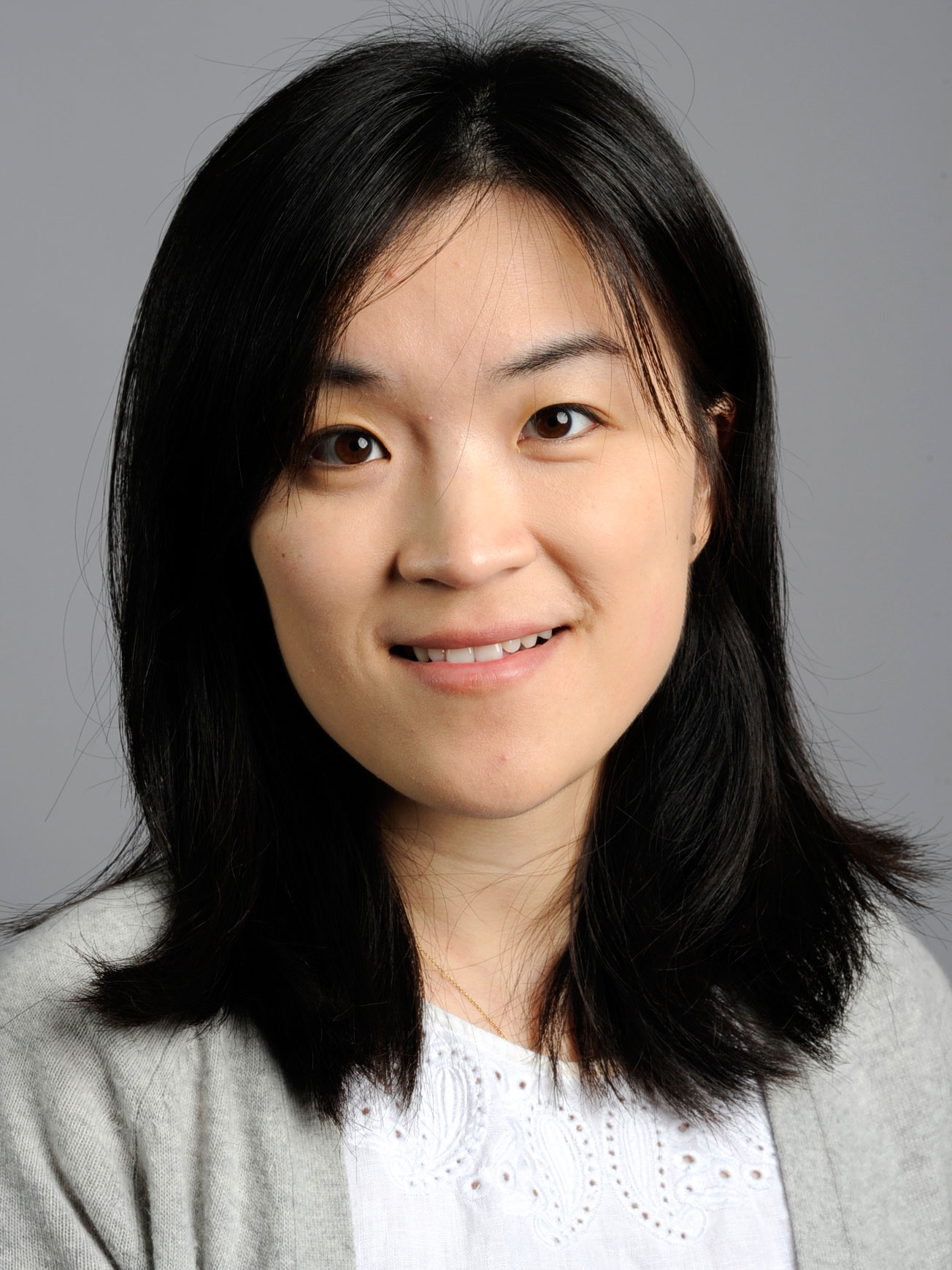}}]{Xiaoning Jin} (M’12) received the B.S. degree in Industrial Engineering from Shanghai Jiao Tong University, Shanghai, China, in 2006, and the M.S. and Ph.D. degrees in Industrial and Operations Engineering from the University of Michigan, Ann Arbor, MI, USA, in 2008 and 2012, respectively. She was an assistant research scientist in mechanical engineering at the University of Michigan until 2016. She is currently an Associate Professor with the Department of Mechanical and Industrial Engineering at Northeastern University, Boston, MA, USA. Her research interests include intelligent manufacturing systems, data analytics and machine learning, diagnostics and prognostics, and decision support tools.
\end{IEEEbiography}

\end{document}